\documentclass[range]{ar2e}

\usepackage{cite}
\usepackage{hyperref}
\usepackage{graphicx}

\begin{document}

\input{epsf.tex}   


\input{psfig.sty}

\bibliographystyle{arnuke_revised}

\jname{Annu. Rev. Nucl. Part. Sci.}
\jyear{2013,}
\jvol{Vol. 63.}
\ARinfo{doi: 10.1146/annurev-nucl-102212-170627}

\newcommand{\mup}{\ensuremath{\mu \mathrm{p} }}
\newcommand{\Ryd}{\ensuremath{R_{\infty}}}
\newcommand{\rp}{\ensuremath{r_{\mathrm p}}}
\newcommand{\Htwo}{\ensuremath{\mathrm H_{2}}}
\newcommand{\Water}{\ensuremath{\mathrm H_{2}\mathrm O}}
\newcommand{\mus}{\ensuremath{\mu \mathrm{s} }}
\newcommand{\mum}{\ensuremath{\mu \mathrm{m} }}
\newcommand{\Ka}{\ensuremath{\mathrm{K}_\alpha }}

\title{Muonic hydrogen and the proton radius puzzle}

\markboth{Proton Radius Puzzle}{Proton Radius Puzzle}

\author{
Randolf Pohl
\affiliation{Max-Planck-Institut f\"{u}r Quantenoptik, 85748 Garching, Germany}
Ronald Gilman
\affiliation{Department of Physics \& Astronomy, Rutgers University, Piscataway, NJ 08854-8019, USA} 
Gerald A.\ Miller
\affiliation{Department of Physics, Univ. of Washington, Seattle, WA 98195-1560, USA}
Krzysztof Pachucki
\affiliation{Faculty of Physics, University of Warsaw,
             Ho\.{z}a 69, 00-681 Warsaw, Poland}
}


\begin{abstract}
  The extremely precise extraction of the proton radius by Pohl {\it et al.}
  from the measured energy difference between the 2P and 2S states of muonic
  hydrogen disagrees significantly with that extracted from electronic
  hydrogen or elastic electron-proton scattering.
  This is the proton radius puzzle.  The origins of the puzzle and
  the reasons for believing it to be very significant are explained. Various
  possible solutions of the puzzle are identified, and future work needed to
  resolve the puzzle is discussed.
\end{abstract}

\maketitle

\section{Introduction}
%
%
\newcommand{\nn}{\nonumber\\&&}
\newcommand{\NNN}{$N\!N\!N$}
\newcommand{\NNNNNN}{$N\!N\!N\rightarrow N\!N\!N$}
\newcommand{\bk}{\bar{k}}
\newcommand{\bq}{\bar{q}}
\newcommand{\bp}{\bar{p}}
\newcommand{\bu}{\bar{u}}
\newcommand{\tPhi}{\tilde{\Phi}}
\newcommand{\tphi}{\tilde{\phi}}
\newcommand{\tX}{\tilde{X}}
\newcommand{\tY}{\tilde{Y}}
\newcommand{\tT}{\tilde{T}}
\newcommand{\tv}{\tilde{v}}
\newcommand{\ttt}{\tilde{t}}
\newcommand{\td}{\tilde{d}}
\newcommand{\tj}{\tilde{j}}
\newcommand{\tG}{\tilde{G}}
\newcommand{\la}{\langle}
\newcommand{\ra}{\rangle}
\newcommand{\bfr}{{\bf r}}
\newcommand{\kslash}{\not\hspace{-0.7mm}k}\newcommand{\lslash}{\not\hspace{-0.7mm}l}
\newcommand{\pslash}{\not\hspace{-0.7mm}p}
\newcommand{\sslash}{\not\hspace{-0.7mm}s}

\newcommand{\qslash}{\not\hspace{-0.7mm}q}
\newcommand{\pslashon}{\not\hspace{-0.7mm}(p-q)_{on}}
\newcommand{\pslashoff}{\not\hspace{-0.7mm}p^{\rm off}}
\newcommand{\munu}{{\mu\nu}} \newcommand{\mn}{{\mu\nu}}
\newcommand{\ben}{\begin{displaymath}}
\newcommand{\een}{\end{displaymath}}
\newcommand{\be}{\begin{equation}}
\newcommand{\ee}{\end{equation}}
\newcommand{\bea}{\begin{eqnarray}}
\newcommand{\eea}{\end{eqnarray}}
\newcommand{\eqn}[1]{\label{#1}}
\newcommand{\eq}[1]{Eq.~(\ref{#1})}
\newcommand{\eqs}[1]{Eqs.~(\ref{#1})}
\newcommand{\fign}[1]{\label{#1}}
\newcommand{\fig}[1]{Fig.~\ref{#1}}
\newcommand{\bPsi}{\bar{\Psi}}
\newcommand{\bPhi}{\bar{\Phi}}
\newcommand{\bphi}{\bar{\phi}}
\newcommand{\bfp}{{\bf p}}
\newcommand{\bfP}{{\bf P}}
\newcommand{\bfq}{{\bf q}}
\newcommand{\bfk}{{\bf k}}                                                           
\newcommand {\bfxi}{\mbox{\boldmath$\xi$}}
\newcommand {\boldsigma}{\mbox{\boldmath$\sigma$}}
\newcommand {\boldgamma}{\mbox{\boldmath$\gamma$}}
\newcommand {\boldxi}{\mbox{\boldmath$\xi$}}
\newcommand {\boldSigma}{\mbox{\boldmath$\Sigma$}}
\newcommand{\suplus} {^{(+)}}
\newcommand{\suminus}{^{(-)}}
\newcommand{\pp}{p^0+i\epsilon}

The recent determination of the proton radius using the measurement of the
Lamb shift in the muonic hydrogen
atom~\cite{Pohl:2010:Nature_mup1,Antognini:2013:Science_mup2} startled the
physics world. The obtained value of 0.84087(39)\,fm differs by about 4\% or 7
standard deviations from the CODATA~\cite{Mohr:2012:CODATA10} value of
0.8775(51)\,fm. The latter is composed from the electronic hydrogenate atom
value of 0.8758(77)\,fm and from a similar value with larger uncertainties
determined by electron scattering~\cite{Bernauer:2010:NewMainz}. The preceding
sentence brings up a number of simple questions. The most prominent are: how
can atomic physics be used to measure a fundamental property of a so-called
elementary particle, why should muonic hydrogen be more sensitive to this
quantity than electronic hydrogen, and why should a 4\% difference between
proton radii extracted using different techniques be important? In the present
Introduction we sketch brief answers. The remainder of the article is devoted
to detailed answers to these and related questions and implications.

The sensitivity of atomic energy levels to the non-zero size of the proton is
determined by the probability that the bound lepton be within the volume of
the proton. This probability is roughly given by the ratio of proton to atomic
volumes: $(r_p/a_B)^3=(\alpha m_r r_p)^3$, where $r_p$ is the proton radius
and $m_r$ is the lepton reduced mass. The muon mass is about 200 times the
electron mass so the muon is about 8 million more times likely to be inside
the proton than the electron.

To be a bit more precise we need to define the proton radius. The mean-square
value $r_p^2$ of the radius is given by
\bea
r_p^2\equiv -6 {dG_E\over dQ^2}\Big\vert_{Q^2=0} ,
\eea
where $G_E$ is the Sachs electric form factor of the proton and $Q^2$ is the
negative of the square of the four-momentum transfer to the proton. We note in
passing that, because of recoil effects, this radius is not actually the
integral of $r^2$ times a true
density~\cite{Miller:2007uy,Miller:2010nz}. However it is $r_p^2$ that is
observable in energy levels of muon and electron hydrogen atoms.

For non-relativistic systems the lepton-proton Coulomb interaction is modified
away from the point Coulomb interaction by the factor $G_E(Q^2)$ which is the
probability amplitude for the proton to absorb the exchanged
photon~\cite{Eides:2006:Book,Bawin:1999ks}. The resulting difference between
the true potential and the potential for a point-like ($G_E=1)$ proton is
given by
\bea 
\delta V(\bfr)\equiv V_C(\bfr)-V_C^{\rm pt}(\bfr)=-4\pi \alpha\int {d^3q\over(2\pi)^3}{(G_E(\bfq^2)-1)\over \bfq^2} e^{-i \,{\bf q}\cdot {\bf r}}.
\label{db}
\eea
Using the expression $G_E(\bfq^2)-1\approx -\bfq^2r_p^2/6$ in \eq{db} is a
very accurate approximation because in atomic physics $ r_p\;q\sim
r_p/a_B\sim10^{-5}$. If this approximation is used, the perturbation $\delta
V$ is a Dirac-delta function at the origin multiplied by a factor that
includes $r_p$, and the resulting energy shift for atomic S-states is given
by~\cite{Karplus:1952zza}:
\bea 
\Delta E=\langle \Psi_S\vert \delta V\vert \Psi_s\rangle ={2\over 3}\pi \alpha \left\vert \Psi_S(0)\right\vert^2r_p^2.
\label{de}
\eea 
The S$_{1/2}$ and P$_{1/2}$ states are degenerate for solutions of both the
Schroedinger and Dirac equations. But with the effects of non-zero proton
extent included, the S$_{1/2}$ state moves up and the P$_{1/2}$ state is not
affected. This gives a contribution to the energy difference between S$_{1/2}$
and P$_{1/2}$ states, part of the Lamb shift.

The difference between the electronic and muonic determinations of the proton
radius is very puzzling.  One possibility is that experimental results are
wrong. But multiple independent electron-proton experiments agree, and the
muonic hydrogen experiment looks more convincing than any of the
electron-proton experiments.  Similarly, many measurements of transition
frequencies in hydrogen are in agreement. As detailed in
Sect.~\ref{sec:H_exp}, the muonic hydrogen charge radius would require a
change of the Rydberg constant~\cite{Mohr:2012:CODATA10} by 6.6 standard
deviations~\cite{Pohl:2010:Nature_mup1,Antognini:2013:Science_mup2}, in
disagreement with the value obtained from many measurements in hydrogen.
Another way out would be to assert that the various QED calculations that give
contributions to the Lamb shift (other than the proton extent) are not
sufficiently accurate. But all of the important contributions have been
obtained by more than one group~\cite{Eides:2006:Book} and checked and
rechecked. Another possibility is that the muon and electron have different
interactions with the proton. But this violates a well-known and well-tested
principle called lepton universality. The failure to understand why the proton
radius can be different in muonic and electronic hydrogen is indeed a puzzle.

\section{Electronic measurements}
\begin{figure}[t]
\begin{center}
\includegraphics[width = 1.0\columnwidth]{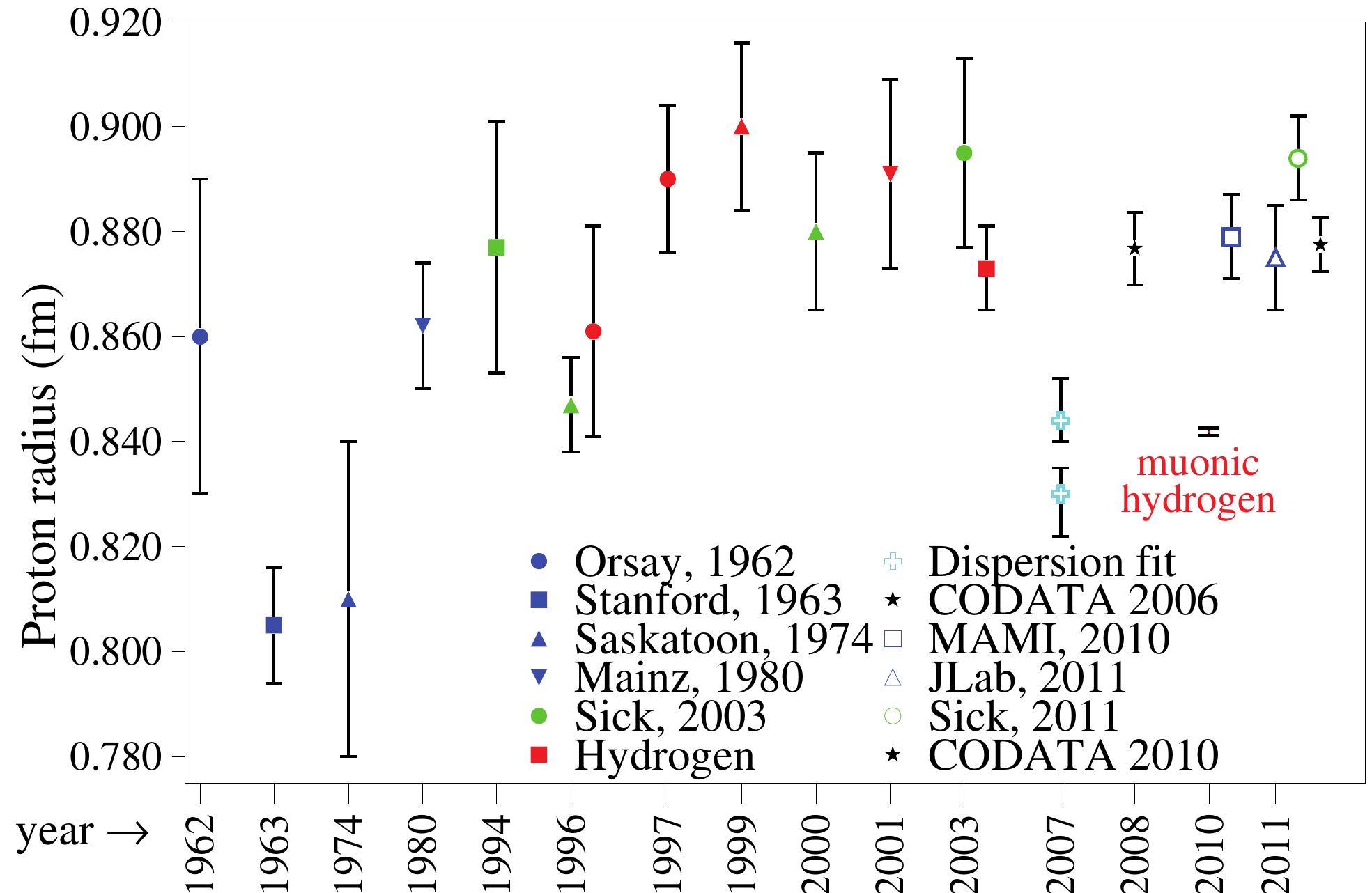}
\caption{Proton radius determinations over time. Electronic measurements
  seem to settle around \rp=0.88\,fm, whereas the muonic hydrogen
  value\cite{Pohl:2010:Nature_mup1,Antognini:2013:Science_mup2} is at 0.84\,fm.
  Values are (from left to
  right):
  Orsay~\cite{Lehmann:1962:RP_Orsay},
  Stanford~\cite{Hand:1963:RP_Stanford},
  Saskatoon~\cite{Murphy:1974:RP_Saskatoon,Murphy:1974:RP_Saskatoon_erratum},
  Mainz~\cite{Simon:1980:RP_Mainz} (all in blue) 
  are early electron scattering measurements.
  Recent new scattering measurements are from 
  MAMI~\cite{Bernauer:2010:NewMainz} and 
  Jlab~\cite{Zhan:2011:JLab_Rp}.
  The green and cyan points denote various reanalyses of the world electron 
  scattering data~\cite{Wong:1994:Rp,Mergell:1995bf,Rosenfelder:1999cd,
    Sick:2003:RP,Belushkin:2007:NuclFF,Sick:2011:Troubles}.
  The red symbols originate from laser spectroscopy of 
  atomic hydrogen and advances in hydrogen QED theory (see 
  \cite{Mohr:2012:CODATA10} and references therein).
  The green and red points in the year 2003 denote the reanalysis of the world 
  electron scattering data~\cite{Sick:2003:RP} and the world data from hydrogen
  and deuterium spectroscopy which have determined the value of \rp{} in the 
  CODATA adjustments~\cite{Mohr:2008:CODAT06,Mohr:2012:CODATA10}
  since the 2002 edition.
  %
}
\label{fig1:Rp_vs_t}
\end{center}
\end{figure}
 
Fig.~\ref{fig1:Rp_vs_t} shows several determinations of the proton charge
radius. Early elastic electron-proton scattering measurements from
Orsay~\cite{Lehmann:1962:RP_Orsay}, Stanford~\cite{Hand:1963:RP_Stanford},
Saskatoon~\cite{Murphy:1974:RP_Saskatoon,Murphy:1974:RP_Saskatoon_erratum} and
Mainz~\cite{Simon:1980:RP_Mainz}, and the various re-analyses of these world
data~\cite{Sick:2003:RP,Blunden:2005:RP}.  Spectroscopy of atomic hydrogen
became sensitive to \rp{} on the percent level in the mid-1990s.

%
%

\subsection{Hydrogen spectroscopy}
\label{sec:H_exp}

Spectroscopy of atomic hydrogen (H) and deuterium (D) has been important the
development of modern physics for more than a century. It was the discovery of
the Lamb shift in H~\cite{Lamb:1947:FShyd} which first showed effects of
quantum electrodynamics (QED) which go beyond the Dirac equation.
Today, the 1S-2S transition in H has been measured with an accuracy of 4 parts
in $10^{15}$~\cite{Parthey:2011:PRL_H1S2S}. Other transitions, most notably
the two-photon-transitions between the metastable 2S state and the states
8S,D~\cite{Beauvoir:1997:H2S8SD} or 12D~\cite{Schwob:1999:Hydr2S12D}, have
been measured with accuracies around 1 part in $10^{11}$.
For a review of the relevant
transition frequencies in H and D see~\cite{Mohr:2012:CODATA10}.

QED describes the energy levels of H with extraordinary accuracy. The test of
QED using measured transition frequencies in H is limited by two of the input
parameters required in the QED calculations, namely the Rydberg constant
\Ryd{} and the rms proton charge radius \rp{}. Hence, one can either supply any
of these two numbers from a source other then H spectroscopy (such as \rp{}
from elastic e-p scattering of muonic hydrogen) and then test the correctness
of QED, or use QED to extract the fundamental constants \Ryd{} and \rp{}.


Somewhat simplified, the energies of S-states in H are given by
\begin{equation}
\label{eq:E_simple}
E(nS) \simeq - \frac{\Ryd}{n^2} + \frac{L_{1S}}{n^3}
\end{equation}
where $n$ is the principal quantum number, and $L_{1S}$ denotes the Lamb shift
of the 1S ground state which is given by QED and contains the effect of the
proton charge radius \rp{}. Numerically, $L_{1S} \simeq ( 8172 + 1.56\,
\rp^2 )$\,MHz when \rp{} is expressed in fm, so the finite size effect on the 1S
level in H is about 1.2\,MHz.

\begin{figure}[t]
\begin{center}
\includegraphics[width = 1.0\columnwidth]{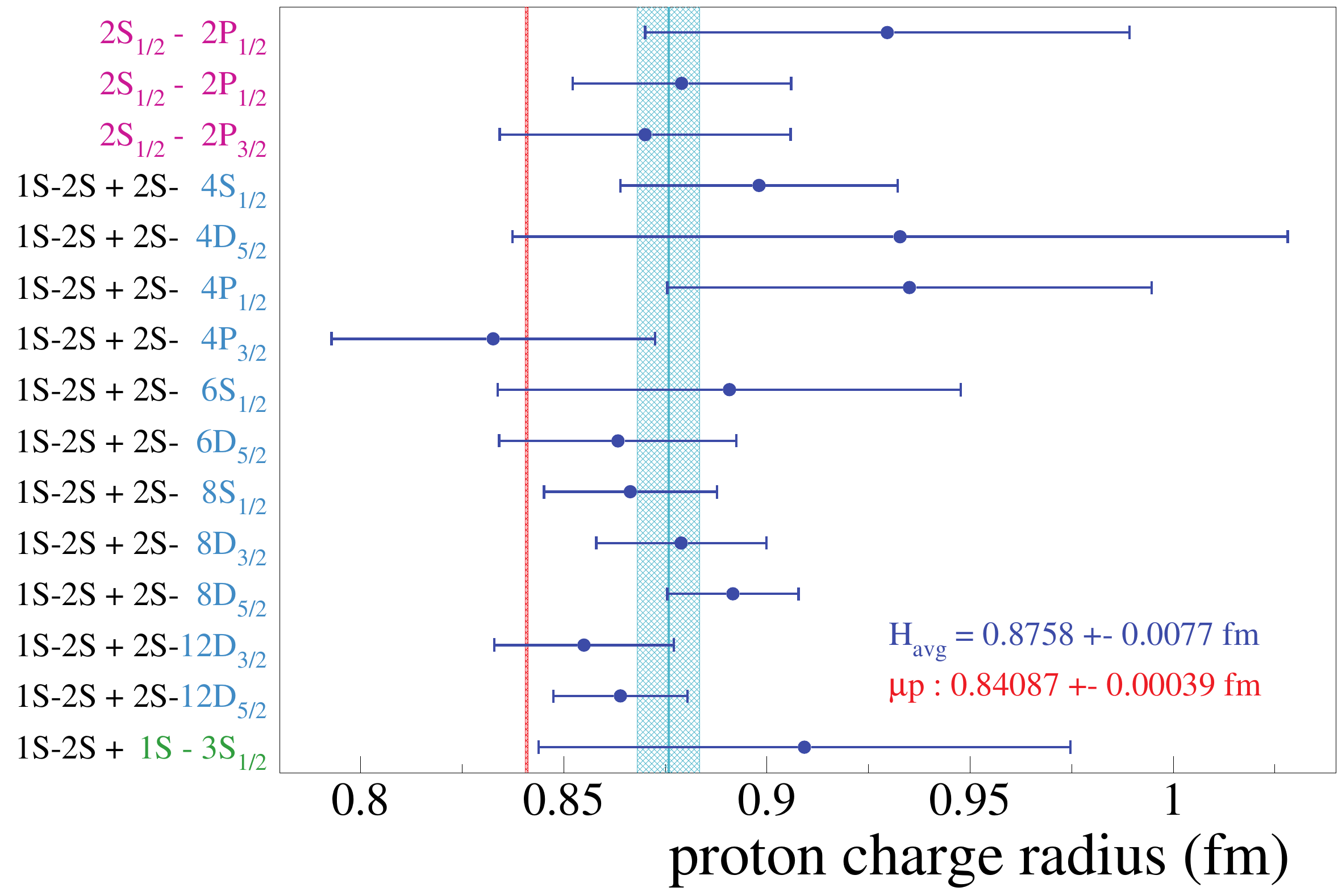}
\caption{Proton charge radii \rp{} obtained from hydrogen
  spectroscopy. According to Eq.~(\ref{eq:E_simple}), \rp{} can best be
  extracted from a combination of the
  1S-2S transition frequency~\cite{Parthey:2011:PRL_H1S2S} and one of
  the 2S-8S,D or 12D
  transitions~\cite{Beauvoir:1997:H2S8SD,Schwob:1999:Hydr2S12D}.
  The value from muonic
  hydrogen~\cite{Pohl:2010:Nature_mup1,Antognini:2013:Science_mup2} is shown with its error bar.}
\label{fig2:Rp_from_H}
\end{center}
\end{figure}

The different $n$-dependence of the two terms in Eq.~(\ref{eq:E_simple})
permits the determination of both \Ryd{} and \rp{} from at least two
transition frequencies in H.
Ideally, one uses the most accurately measured 1S-2S
transition~\cite{Parthey:2011:PRL_H1S2S} and one of the 2S-8S,D/12D
transitions~\cite{Beauvoir:1997:H2S8SD,Schwob:1999:Hydr2S12D}.  
The former contains the maximal 1S Lamb shift $L_{1S}$ and is hence maximally
sensitive to \rp{}. The latter contain only smaller Lamb shift contributions
due to the $1/n^3$ scaling in Eq.~(\ref{eq:E_simple}), and hence determine
\Ryd{}. Fig.~\ref{fig2:Rp_from_H} shows the difference values of \rp{} obtained
by combining the 1S-2S transition and each of the other precisely measured
transitions in H. In addition, it contains three values of \rp{} obtained from
a direct measurement of the 2S-2P transitions in H. These are not sensitive to
the Rydberg constant.

From Fig.~\ref{fig2:Rp_from_H} one can observe that all \rp{} values from H
favor a larger \rp{} around 0.88\,fm. Still, half of the individual \rp{}
values agree with the muonic hydrogen value of 0.84\,fm on the level of
$1\sigma$. In fact, only the 2S-8D$_{5/2}$
transition~\cite{Beauvoir:1997:H2S8SD} disagrees with the muonic \rp{} value
on the level of $3\sigma$.
The discrepancy between the combined value from H, as obtained in the
elaborate CODATA adjustment of the fundamental
constants~\cite{Mohr:2012:CODATA10}, and the muonic hydrogen value, is about
$4.4\sigma$.


%
%

\subsection{Elastic electron-proton scattering}

Elastic electron scattering has been used to measure the
electromagnetic structure of nucleons and nuclei for
about six decades, and been reviewed for nearly as 
long \cite{Perdrisat:2006hj,Arrington:2011kb}.
For the proton charge and magnetic radii, the key points are that the structure is 
encoded in two momentum-space, relativistic invariants
-- the electric and magnetic form factors $G_E(Q^2)$ and $G_M(Q^2)$ -- 
and the radii are determined from the slope of the form factors at 
four-momentum transfer $Q^2$ = 0.
(In this section the topic of the proton magnetic radius arises at
times, but usually our remarks will be directed towards the charge radius.)
In the one-photon exchange approximation, the measured
experimental cross section is related to the form factors
by \begin{equation}
{{d\sigma}\over{d\Omega}}_{exp} = 
{{d\sigma}\over{d\Omega}}_{point} \times \left [
{{G_E^2(Q^2) + \tau G_M^2(Q^2)} \over {1+\tau}} + 2 \tau
\tan^2(\theta/2)  G_M^2(Q^2) \right].
\label{eq:epscatt}
\end{equation}
Here $\tau = Q^2/4m^2$ and $\theta$ is the lab scattering angle.
Traditionally, cross section measurements were corrected
for certain ``radiative corrections'', higher order diagrams
beyond the one-photon exchange approximation, not included in 
Eq.~\ref{eq:epscatt}.
Form factors were extracted using Rosenbluth separations, multiple
cross section measurements at different beam energies and scattering
angles, resulting in the same momentum transfer.
Then fits of the form factors with simple, sometimes
theoretically inspired, parameterizations were 
used to determine the radii.
However, due to a variety of issues with experiments and 
corrections, early radii extractions in the literature should no
longer be taken seriously. One should instead be concerned
with analyses of the past several years that pay attention
to the issues mentioned below and are careful about extracting 
the radius.

Many of the issues in extracting a proton radius from scattering
data have been known for over a decade
\cite{Rosenfelder:1999cd,Sick:2003:RP}. These include
\begin{itemize}
\item The treatment of systematic uncertainties in many experiments was
often unclear, and apparently optimistic. Fits to world data, taking
the uncertainties as reported, do not give reduced $\chi^2$
sufficiently close to 1.
\item The radius determination is mainly sensitive to low $Q^2$ data,
and world data fits can be satisfactory overall without going through
the low $Q^2$ data. 
\item Fits extracting the radius using a single functional form
typically ignored the uncertainty from model dependence.
\item In principle the Taylor series expansion $G_E = 1 - Q^2 r_p^2 / 6 + Q^4
  r_p^4 / 120 \ldots$ gives a model independent radius determination.
  In practice the finite range of and uncertainties in the data lead to correlations
  between the coefficients $r_p^n$ and possible truncation
  errors. Because the coefficients $r_p^n$ grow with order, there is
  no $Q^2$ range where one term sufficiently dominates that its factor
  $r_p^n$ can be fixed and used in fitting lower order terms to lower $Q^2$ data.
\item Issues with the absolute normalization affect the extracted radius,
  but floating the normalization decreases sensitivity to the radius, and
  increases sensitivity to possible $Q^2$ dependent errors - the rapid
  change of the cross section and data rates with $Q^2$ is a problem.
\item Early radius extractions lacked Coulomb corrections,
the acceleration of the electron in the proton's Coulomb field,
which tend to increase the extracted radius about 0.01 fm \cite{Rosenfelder:1999cd}.
\end{itemize}

The observation of differences in the ratio $G_E(Q^2)/G_M(Q^2)$
measured by Rosenbluth techniques and by polarization transfer
techniques, first observed in \cite{Jones:1999rz}
ultimately led to
the recognition of the importance of hard two-photon exchange
corrections~\cite{Arrington:2011dn}, ignored in earlier work. Calculations of 
two-photon exchange depend on the off-shell structure of the
proton, and are not under precise theoretical control.
Experimental measurements generally constrained two-photon
effects to be no more than $\approx$1\%.
This topic continues to be
under active investigation, due to its potential impact on knowledge
of the proton structure, and of the radius.
High precision experiments are under way to determine
observables that depend on two-photon exchange, such as difference
in $e^+p$ and $e^-p$ cross sections, single spin asymmetries,
and variations in form factor ratios extracted at the same momentum transfer
as a function of the beam energy or scattering angle.
While two-photon exchange should be under better control in the
near future, it is currently believed to be a small correction 
for low $Q^2$ data of most importance to the radius puzzle, and
mainly of importance in determining the magnetic radius \cite{Arrington:2012dq}.

A renewed interest in the long-range, low-$Q^2$ structure of the proton led to
several new electron scattering experiments performed at the same time as the
muonic hydrogen experiment. 
The Bates BLAST collaboration \cite{Crawford:2006rz} used a polarized
electron beam incident on a polarized gas target to determine
that the form factor ratio $\mu G_E/G_M$ was close to unity
for $Q^2$ $\approx$ 0.15 $\to$ 0.6 GeV$^2$.
The Jefferson Lab LEDEX collaboration \cite{Ron:2007vr,Ron:2011rd}
used polarization transfer to determine that the form factor ratio 
deviated significantly from unity by 0.3 - 0.4 GeV$^2$, subsequently
confirmed and improved upon in \cite{Zhan:2011:JLab_Rp}.
Note that while the polarization data give a form factor ratio, which does not
determine the radius, the ratio does help constrain normalizations of
cross section data during fits, leading to an improved value of the radius.
The analysis of \cite{Zhan:2011:JLab_Rp} gives confirms the larger proton
radius of electron measurements, $r_p$ = 0.875 $\pm$ 0.008$_{exp}$
$\pm$ 0.006$_{fit}$.
Data that extend the form factor ratios to even lower
$Q^2$, 0.01 GeV$^2$ - 0.06 GeV$^2$ are currently under
analysis \cite{Ron:2011zz}.

Most importantly for the radius puzzle, the Mainz A1 collaboration
\cite{Bernauer:2010:NewMainz} measured 1422 precise relative cross sections
over a wide range of angles (20$^{\circ}$ - 135$^{\circ}$) and 
beam energies (180 MeV - 855 MeV), corresponding to
$Q^2$ = 0.0038 $\to$ 0.98 GeV$^2$.
Experimental systematic uncertainties were controlled by using one spectrometer 
as a luminosity monitor, and by determining cross sections with two 
spectrometers moved through multiple, overlapping angle settings.
The data were fit with a variety of functional forms using 31
normalization constants.
Generally it was found that a good $\chi^2$/d.o.f.,
$\approx$1.14, could be obtained
with flexible forms (polynomials or splines) but not with inflexible
forms, such as dipoles.
Polynomials and splines led to slightly different extracted
radii, leading to $r_p$ = 0.879 $\pm$ 0.005$_{stat}$
$\pm$ 0.004$_{syst}$ $\pm$ 0.002$_{model}$ $\pm$ 0.004$_{group}$,
where the final uncertainty comes from the polynomial vs.\ spline difference.

The fits of \cite{Bernauer:2010:NewMainz} and \cite{Zhan:2011:JLab_Rp} 
use entirely independent data sets, and give form factor ratios 
in good agreement in the more limited $Q^2$
range of the polarization data \cite{Zhan:2011:JLab_Rp}.
The proton charge radii from the two fits also agree, but
the magnetic radii are significantly different.
A large part of the difference appears to result from different
treatments of the two-photon exchange corrections
\cite{Arrington:2011kv,Bernauer:2011zza}, and it appears
that the difference in magnetic radii has no effect on the
determination of the electric radius.

These new experimental reports reinforce the proton radius puzzle.
Several independent analyses of the proton radius have also been made.
Here we highlight four recent contradictory results.

Dispersion relations and the vector meson dominance
model have long been used to fit the form factors,
with the space-like form factors resulting from poles
in the time-like region. Various other features
such as continuum contributions and expected
high-$Q^2$ asymptotic behavior have been added.
The most recent work of this sort \cite{Lorenz:2012tm}
finds $r_p$ = 0.84 $\pm$ 0.01 fm with $\chi^2$/d.o.f $\approx$ 2.2.
Several notes are in order.
Neutron and proton data are simultaneously fit.
The Mainz A1 data set normalization is allowed to float, and in most
cases the change is less than 1\%.
Statistically, the reduced $\chi^2$ of 2.2 indicates that the data are not well described by the fit; while in general
this could result from underestimated uncertainties in the data,
the lower $\chi^2$ of the flexible fits of \cite{Bernauer:2010:PhD,Bernauer:2010:NewMainz}
suggest this is not the case here.
However, visually the fits of \cite{Lorenz:2012tm} go through
all the lower $Q^2$ data well, and start to diverge somewhat
at larger values of $Q^2$.
This suggests that for determining the radius the large value of $\chi^2$ 
is not an issue, but that there may be an issue with inappropriate
strength in higher order terms -- thinking in terms of the Taylor
series expansion -- leading to an incorrect result for the radius
term.
The smaller charge radius in these dispersion relation fits is in line with
previous fits of this type \cite{Belushkin:2006qa,Mergell:1995bf}.

The $z$ expansion -- see \cite{Hill:2010yb} for its application to the
proton form factors -- provides a method of incorporating
physical constraints (analyticity) in choosing a functional form to
fit the form factor data.
The form factor is parameterized as a power series in 
a complex variable $z(Q^2)$ that is constrained to lie within the
unit circle.
The power series expansion of the form factor is model independent,
unlike the particular functional forms chosen in the dispersion
relations analyses of
\cite{Lorenz:2012tm,Belushkin:2006qa,Mergell:1995bf} and others.
The benefits of this transformation include a maximum value of
$z$ and coefficients that are bounded,
guaranteeing a finite number of terms will be adequate.
A fit with only the proton data yields 
$r_p$ = 0.870 fm $\pm$ 0.023 fm $\pm$ 0.012 fm;
while including neutron data and the $\pi\pi$ continuum leads to 
$r_p$ = 0.871 fm $\pm$ 0.009 fm $\pm$ 0.002 fm $\pm$ 0.002 fm -
see \cite{Hill:2010yb} for details.
While the analysis continues to suggest a larger proton radius, it
also suggests that the proton radius uncertainties are underestimated 
by other recent analyses.

Some of the issues discussed previously in relation to fits were
revisited in recent work by Sick \cite{Sick:2011zz,Sick:2012zz}.
The extraction of the proton radius was found to be most sensitive
to data in the range 0.01 GeV$^2$ $\to$ 0.06 GeV$^2$ (0.5 fm$^{-1}$
$\to$ 1.3 fm$^{-1}$).
Fit functions with polynomials in a denominator can, with negative
terms in the denominator, generate pathological behaviors -- poles
in the form factors at high $Q^2$ outside the range of data.
Thinking in terms of the Fourier transform of a spatial distribution, 
this generates non-physical oscillatory behavior that extends to 
large $r$ and affects the radius extracted.
Despite the low $Q^2$ of the Mainz data, fits of the data would be more
reliable if the data extended to lower $Q^2$ and / or did not have a
floating normalization.
Parameterizations that correspond to a more sensible large $r$
fall off should be safer.
A sum-of-Gaussians $r$-space fit gave a stable radius determination, 
without the issues
discussed, of $r_p$ = 0.886 fm $\pm$ 0.008 fm.

Recent unpublished work by C.E.~Carlson and K.~Griffioen points out that a linear 
fit to the lowest $Q^2$ Mainz data, $Q^2$ $<$ 0.02 GeV$^2$, where 
$G_E$ appears to be entirely linear, yields $r_p$ $\approx$ 0.84 fm, 
rather than $\approx$ 0.88 fm. 
But great care must be taken in doing any such fit
\cite{Sick:2003:RP}.
Since for the proton $r_p^4$ $>$ 0, the quadratic term in the Taylor 
series expansion nearly guarantees that a linear fit to low $Q^2$ 
data will underestimate the radius.
An estimate of the potential size of the effect can be made with
the Kelly form factor parameterization \cite{Kelly:2004},
which has $r_p$ $\approx$ 0.86 fm.
A fit to pseudodata up to 0.02 GeV$^2$, with the density and 
statistics of the Mainz data set, yields 
$r_p$ $\approx$ 0.84 fm $\pm$ 0.01 fm.
Adding in a quadratic term leads to uncertainties on $r_p$
too large to distinguish between 0.84 fm and 0.88 fm.
Similar concerns apply to an extension of this approach up to 0.2 GeV$^2$.
An extensive study of this issue has been performed by M.O.~Distler.

To summarize, the apparently simple problem of determining the slope
of the form factor at $Q^2$ = 0 has numerous potential pitfalls.
The weight of the evidence at this point continues to favor a
larger radius, about 0.88 fm, but suggests that claiming an
uncertainty at the 0.01 fm level is optimistic.

\section{Measurement in muonic hydrogen}
%
%
\label{sec:mup_meas}

A measurement of the Lamb shift in muonic hydrogen (\mup{}) was initially
considered half a century ago as a test of electron vacuum polarization
effects~\cite{DiGiacomo:1969:MUP2S2P}, complementary to the Lamb shift in
regular, electronic hydrogen~\cite{Lamb:1947:FShyd} which is dominated by the
electron self-energy~\cite{Bethe:1947:SE}.
The first observation of x-rays from muonic hydrogen succeeded shortly
afterwards~\cite{Placci:1970:XraysMuP}.

Laser spectroscopy of the 2S-2P transition in \mup{} requires, however, \mup{}
atoms in the metastable 2S state. Several groups failed to observe such
long-lived \mup(2S) atoms when muons are stopped in molecular hydrogen
gas~\cite{Anderhub:1977:Search2S,Egan:1981:LongLived2S,Anderhub:1984:KIntRatMup,Boecklin:1982:PhD},
see also~\cite{Pohl:2011:Conf:PSAS2010}.
The first observation of long-lived \mup{} atoms in the 2S
state~\cite{Kottmann:1999:EXATconfProc,Pohl:2001:MolecQuenchMup,Pohl:2006:MupLL2S}
was the starting point~\cite{Taqqu:1999:SpectrLSmup,Pohl:2000:ExpMeasMup,Kottmann:2001:Conf:Trieste} of the recent Lamb shift measurement~\cite{Pohl:2010:Nature_mup1,Antognini:2013:Science_mup2}.

When negative muons are stopped in molecular \Htwo{} gas at low pressure
(1\,mbar at room temperature), about 1\% of the muons form \mup(2S)
atoms~\cite{Kottmann:1999:EXATconfProc,Pohl:2009:EXA08} with a lifetime of
about 1\,\mus~\cite{Pohl:2006:MupLL2S}. At higher gas pressures the 2S state
is quickly de-excited in collisions with \Htwo{} molecules.

A novel beam line for low-energy negative
muons~\cite{Kottmann:2001:Conf:Trieste} was built at Paul-Scherrer-Institute
(PSI) in Switzerland which delivers about 600 $\mu^-$ per second with a
kinetic energy between 3\,keV and 6\,keV. About half of the muons stop in
1\,mbar \Htwo{} gas within a target vessel which has a length of 20\,cm along
the muon beam axis. The transverse dimensions of the muon beam are $0.5 \times
1.5$\,cm$^2$. Hence the muon stop volume is small enough to be illuminated
with laser light of sufficiently high fluence (see below).
The muons arrive at random times, so each muon has to be detected before it
enters the target vessel. Two stacks of ultra-thin carbon foils, separated by
$\sim 35$\,cm, are used to detect the muon with high efficiency (80\% and
70\%, respectively) using secondary electrons ejected by muons crossing the
carbon foils. A signal in both carbon foil stacks, with the correct
time-of-flight for a low-energy muon traveling the distance between the
carbon stacks, serves as a start signal for the data acquisition and the
pulsed laser system.
The pulsed laser
system~\cite{Antognini:2005:6mumLaser,Antognini:2009:Disklaser} delivers 5\,ns
long pulses of light with a wavelength tunable from 5.5\,\mum{} to
6.1\,\mum{}, with a pulse energy of about 0.25\,mJ. A cw-pumped Yb:YAG disk
laser system~\cite{Giesen:1994:DiskConcept,Antognini:2009:Disklaser} produces
100\,mJ of pulsed pump light only a few hundred nanoseconds after a
muon-induced trigger signal at a random time. This light is used to amplify
red light, tunable around 708\,nm, using Ti:sapphire crystals. The red pulses
are converted to the desired infrared (IR) wavelength at 5.5-6\,\mum{} via
three sequential Stokes shifts in a high-pressure Raman
cell~\cite{Rabinowitz:1986:seqRamanLaser}.
The frequency of the laser light is stabilized and controlled at 708\,nm, and
calibrated in the IR using well-known absorption lines of water vapor
(\Water{}). These \Water{} lines are known with an accuracy of a few
MHz, but pulse-to-pulse energy fluctuations of
the laser system cause a broadening of the laser line width resulting in the
final 300\,MHz uncertainty of the laser frequency
calibration~\cite{Pohl:2010:Nature_mup1}.

A non-resonant multi-pass mirror cavity~\cite{Pohl:2005:MHL} inside the target
gas vessel ensures efficient illumination of the muon stop volume inside the
\Htwo{} target gas. The estimated laser fluence is about 6\,mJ/cm$^2$, which
results in approximately 30\% transition probability when
the laser is tuned to the center of the resonance.

The successful 2S-2P laser excitation is signaled by the emission of a
1.9\,keV \Ka{} x-ray which is emitted in the radiative 2P-1S de-excitation that
follows the 2S-2P transition immediately. The x-rays are detected in 20
large-area avalanche photo diodes (LAAPDs, $14 \times 14$\,mm$^2$ active area
each)~\cite{Fernandes:2003:NIM,Ludhova:2005:LAAPDs} which are mounted close to
the muon stop volume.

The experiment~\cite{Pohl:2010:Nature_mup1,Antognini:2013:Science_mup2} records time spectra of \Ka{}
x-rays for various laser frequencies (i.e.\
wavelengths)~\cite{Nebel:2007:SMH}. For each time spectrum, the number of
\Ka{} x-rays in the time window at which the laser illuminates the muon stop
volume is normalized by the ``prompt'' \Ka{} x-rays which are emitted by the
99\% of the muons which do not populate the metastable 2S state, but proceed
directly to the 1S ground state.

\begin{figure}[h]
\bigskip
\begin{center}
\includegraphics[width = 1.0\columnwidth]{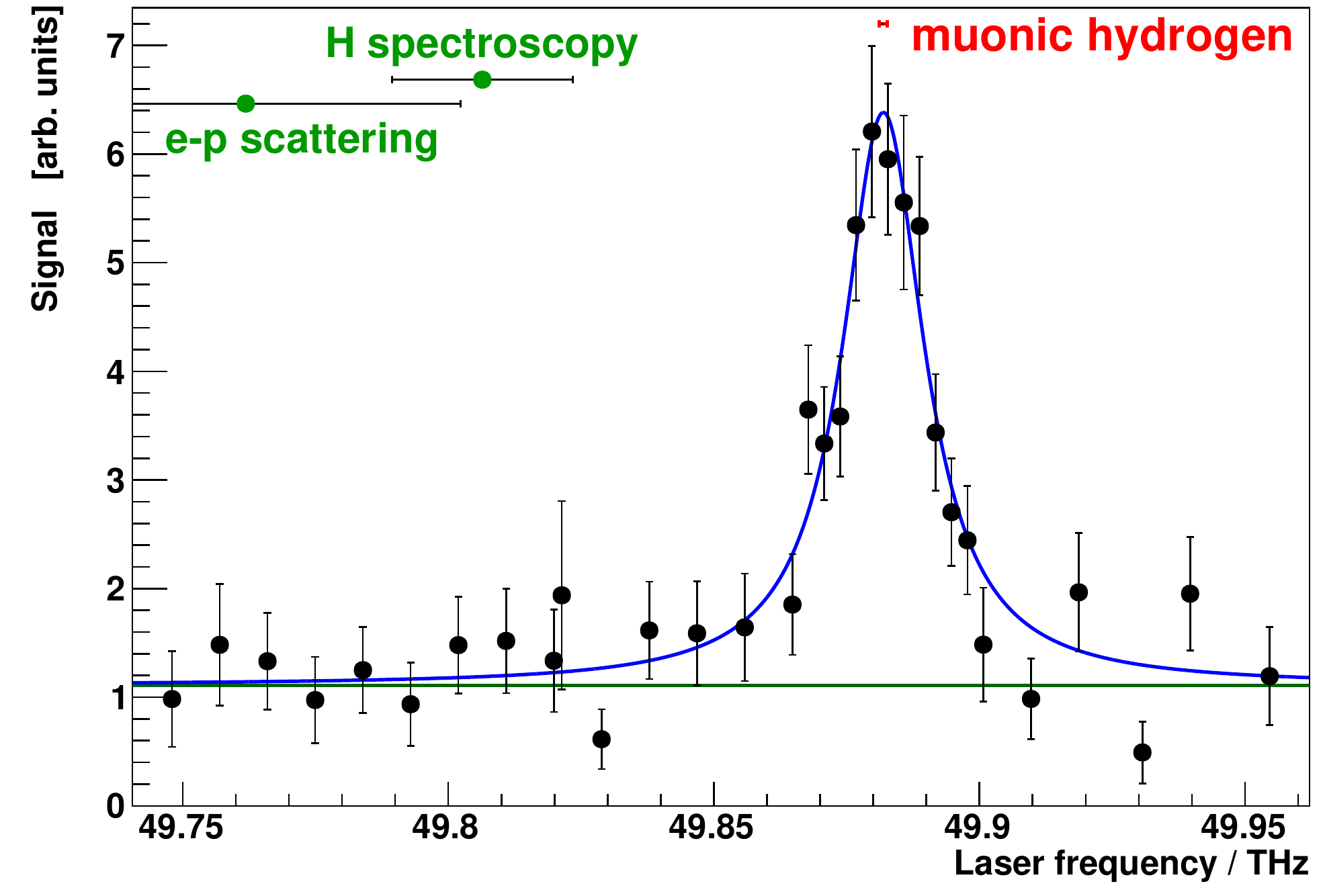}
\caption{Resonance in muonic hydrogen, together with the positions predicted 
using the proton radii from elastic electron-proton scattering using pre-2009
world data~\cite{Sick:2003:RP,Blunden:2005:RP} and the CODATA-2006 value from
H spectroscopy~\cite{Mohr:2008:CODAT06}.
}
\label{fig3:mup_resonance}
\end{center}
\end{figure}

The resonance curve in Fig.~\ref{fig3:mup_resonance} 
is obtained by plotting the normalized number of
laser-induced \Ka{} x-rays as a function of laser frequency. The fit of this
resonance curve reveals the resonance frequency and hence the Lamb shift in
muonic hydrogen~\cite{Pohl:2010:Nature_mup1,Antognini:2013:Science_mup2}.

%

\section{Potential solutions}
The various possible explanations of the proton radius puzzle are discussed in this section.
%
%
\subsection{Muonic hydrogen experiment}

The discrepancy between the observed resonance position in muonic
hydrogen~\cite{Pohl:2010:Nature_mup1,Antognini:2013:Science_mup2} and the predictions using the \rp{}
values from elastic electron scattering~\cite{Sick:2003:RP,Blunden:2005:RP} or
hydrogen spectroscopy~\cite{Mohr:2008:CODAT06} (which together form the CODATA
values 2006 and 2010) is enormous, as can be seen in Fig.~\ref{fig3:mup_resonance}.

The resonance center is 75\,GHz away from the central value of the CODATA
prediction. This corresponds to about four times the width of the resonance.
The statistical uncertainty of the resonance center is about 700\,MHz, i.e.\
1/100 of the discrepancy. The systematic uncertainty of the measurement is
300\,MHz. It is due to the laser frequency calibration mentioned in
Sect.~\ref{sec:mup_meas}. Other systematic effects like ac- and dc-Stark
shifts, Zeeman shifts, Doppler shifts etc.\ have been investigated but were
found to be even smaller~\cite{Pohl:2010:Nature_mup1}.

It is obvious from these numbers that the discrepancy is severe, and it is
difficult to imagine an effect that could shift the resonance position by 4
line widths. Jentschura~\cite{Jentschura:2011:AnnPhys2} found that the
presence of an electron could result in a shift of the resonance position if
the distance between the electron and the \mup(2S) atom was about 1
(electronic) Bohr radius. He suggested that the spectroscopy might have
happened not on a \mup(2S) atom, but on the molecular ion $p \mu e^-$.

Karr and Hilico~\cite{Karr:2012:3body} studied the 3-body systems $p \mu e^-$
and $p p \mu^-$ and found that these are not stable. In addition, the
molecular ion $p p \mu^-$ has been calculated to be very
short-lived~\cite{Froelich:1995:Sidepath,Lindroth:2003:PPMU_with_erratum,Kilic:2004:Decay_ppmu}. The
observation of the quenching of long-lived \mup(2S)
atoms~\cite{Pohl:2006:MupLL2S} supports this picture.  Hence, the proton radius
puzzle cannot be explained by 3-body physics and molecular ion formation.


%

\subsection{Theory of  muonic hydrogen}

Muonic hydrogen, an atom consisting of the proton and the muon
is very similar to  regular ``electronic'' hydrogen, so  the 
theoretical description of the two atoms  has many elements in common. 
The main difference is in the mass of the lepton.
The muon is about 207 times heavier than the electron,
and the muon-proton mass ratio $\eta$, according to
 NIST~\cite{Mohr:2012:CODATA10}, is given by
$\eta = 0.112 609 5272(28)$. This is not very small as is the case for  $e$H, 
and therefore the static nucleus approximation is not necessarily the best 
starting point for the theoretical description of $\mu$H. 
Nevertheless, the traditional nonperturbative approach
introduced by Borie and Rinker in a seminal work \cite{BorieRinker:1982:muAtoms},
continued by Borie \cite{Borie:2005:LSmup,Borie:2012:LS_revisited},  
and very recently by Indelicato \cite{Indelicato:2012} and Carroll {\em et al.} 
\cite{Carroll:2011:NP}
is based on the Dirac equation and corrected for the finite nuclear mass
by including some additional contributions. Here, we  
present an approach, symmetric in the proton and the muon, 
that can also be applied to electronic hydrogen.
There are no significant numerical differences between both approaches~\cite{Antognini:2012:Theory},
but the one presented below, to our opinion, is simpler and more elegant,
since all the corrections can be accounted for in a systematic way.

The binding energy $E$ of a two-body system, according to
quantum electrodynamics (QED) is a function
of the fine structure constant $\alpha$ and of the muon-proton
mass ratio $\eta = m_\mu/m_p$:
\begin{equation}
E = m_\mu {\cal E}(\alpha, \eta),
\end{equation}
where we 
assume that both constituent particles are point-like.
This binding-energy formula also neglects the presence of
electron-positron pairs, which are incorporated through the Uehling
potential discussed below. Even with these simplifications,
there is no single equation
which gives the exact binding energy,
as  is the case for the Dirac or Schr\"odinger equation. 
To some extent, the Bethe-Salpeter equation can be regarded as 
the most general bound state equation,
but the kernel of this integro-differential equation 
can be given only perturbatively in the fine structure constant,
and thus the exact solution is not known. 
In the approach presented here, traditionally called
nonrelativistic quantum electrodynamics (NRQED),
one expands the binding energy ${\cal E}(\alpha, \eta)$
in powers of the fine structure constant $\alpha$, with
$\alpha^{-1} = 137.035 999 074(44)$ and derives an exact closed formula
for the expansion coefficients. In most cases these coefficients 
are expectation values of some effective Hamiltonian, and thus can easily
be calculated.

The description of the QED bound two-body system begins with 
the leading order part, which is the nonrelativistic energy of two particles
interacting via the Coulomb force in the reference frame
with  vanishing total momentum:
\begin{equation}
H_0  =  \frac{p^2}{2\,m_\mu}+  \frac{p^2}{2\,m_p} -\frac{\alpha}{r}.
\end{equation}
According to the Schr\"odinger equation the binding energy 
is a function of the principal quantum number only
\begin{equation}
E_{0}(n) = -\frac{m_r\,\alpha^2}{2\,n^2},
\end{equation}
where $m_r$ stands for the reduced  mass, and $\hbar = c = 1$.
If so, the difference $\Delta E\equiv E_{LS}$ in binding energies
of $2S_{1/2}$ and $2P_{1/2}$ states
\begin{equation}
E_{LS} \equiv E_{2P_{1/2}} - E_{2S_{1/2}}  =0.
\end{equation}
 Departures of $E_{LS}$ from zero, traditionally called the Lamb shift,
 therefore arise from additional corrections to the  energy.

\begin{figure}[!htb]
\begin{center}
\includegraphics[height=0.5\textheight]{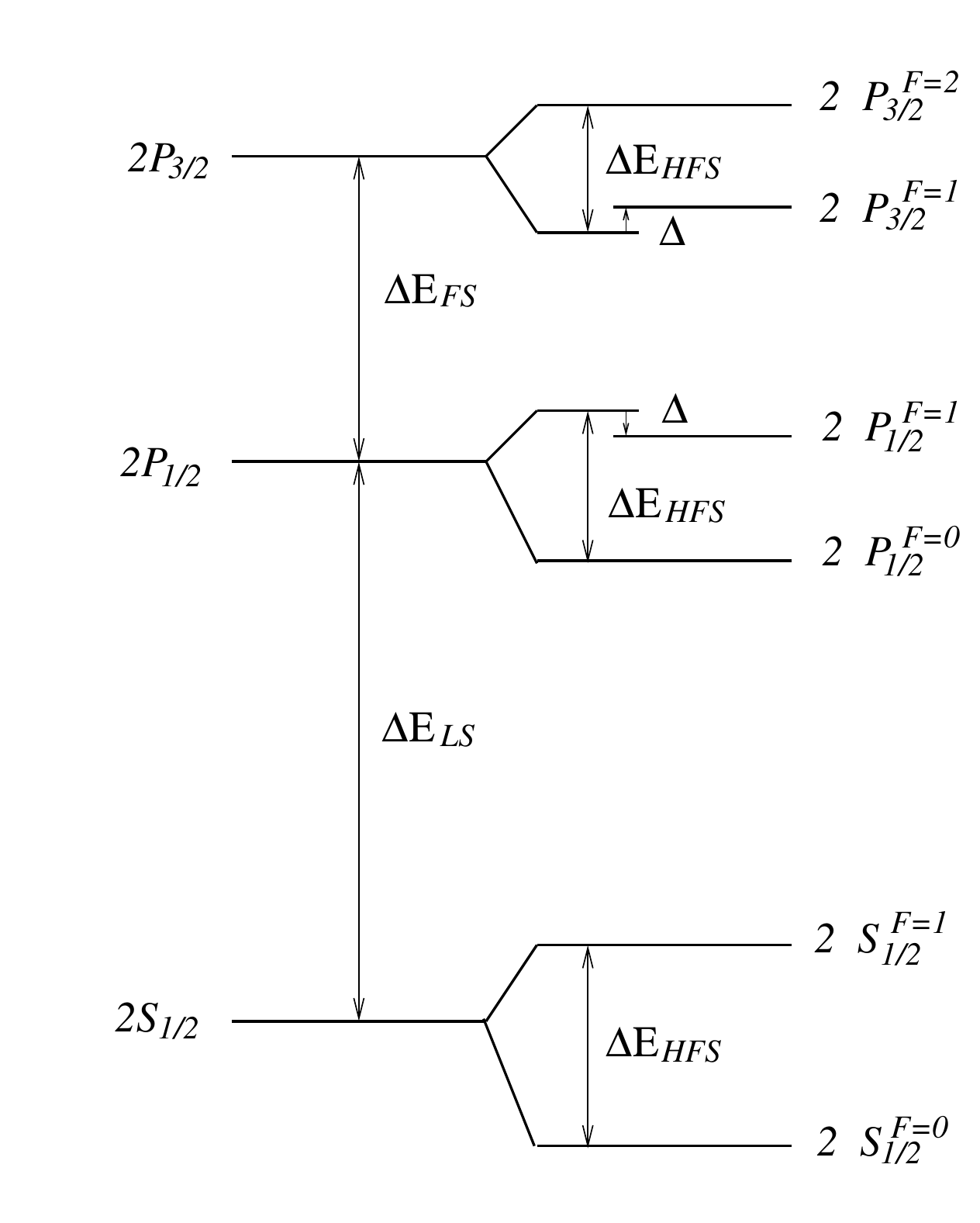}
\caption{Level scheme of muonic hydrogen for $n=2$ shell,
the artificial $2P_{1/2}$ and $2P_{3/2}$ levels corresponds to centroid
for $\Delta = 0$. Numerical values for level splittings are presented in Eq. (\ref{muH30}).}
\label{fig4:muonicH}
\end{center}
\end{figure}

The first set of corrections we discuss involves  
relativistic effects, although these are not the largest ones. 
In the static nucleus 
approximation the binding energy of $2S_{1/2}$ and $2P_{1/2}$ states 
according to the Dirac are also exactly the same.
Moreover, the leading recoil corrections also do cancel,
therefore relativistic effects in muonic hydrogen 
 $\sim\eta^2\,\alpha^4$ and thus are quite small.
How are these relativistic corrections calculated?
According to the NRQED approach, the leading relativistic
corrections can be expressed in terms of the so called Breit-Pauli 
Hamiltonian~\cite{Bethe:1957}
\begin{eqnarray}
H_{BP}  &=& H_0 + \delta H_{BP}\nonumber \\
\delta H_{BP} & = & -\frac{p^4}{8\,m_\mu^3} - \frac{p^4}{8\,m_p^3}
-\frac{\alpha}{2\,m_\mu\,m_p}\,p^i\,
\left(\frac{\delta^{ij}}{r} + 
\frac{r^i\,r^j}{r^3}\right)\,p^j
\\&& 
+ \frac{2\,\pi\,\alpha}{3}\,\biggl(\langle r^2_p \rangle
+\frac{3}{4\,m_\mu^2} + \frac{3}{4\,m_p^2}\biggr)\,\delta^3(r)
\nonumber \\&& 
+\frac{2\,\pi\,\alpha}{3\,m_\mu\,m_p}\,g_\mu\,g_p\,
\vec s_\mu \cdot\vec s_p\,\delta^3(r)
-\frac{\alpha}{4\,m_\mu\,m_p}\,g_\mu\,g_p\,\frac{s_\mu^i\,s_p^j}{r^3}\,
\biggl(\delta^{ij}-3\,\frac{r^i\,r^j}{r^2}\biggr)
\nonumber \\ &&
+\frac{\alpha}{2\,r^3}\,\vec r\times\vec p\,\biggl[
\vec s_\mu \,\biggl(
\frac{g_\mu}{m_\mu\,m_p} 
+\frac{(g_\mu-1)}{m_\mu^2}\biggr)
+\vec s_p \,\biggl(
\frac{g_p}{m_\mu\,m_p} +
\frac{(g_p-1)}{m_p^2} \biggr)\biggr]\,,
\nonumber
\end{eqnarray}
where $\vec s_\mu$, $\vec s_p$ are spin operators of
the muon and the proton, and $g_\mu$, $g_p=5.585\,694\,712(46)$
are $g$-factors. The interaction $\delta H_{BP}$ includes also the effect due to 
the finite proton charge radius $r_p$ together with the Darwin terms
in order to make it clear that these two effects are separated.
$H_{BP}$ was derived years ago before NRQED was formulated,
nevertheless we use it  as a part of the NRQED approach.
The related correction to energy, neglecting 
terms with the proton spin $\vec s_p$,
the proton charge radius $r_p$ and assuming $g_\mu = 2$, 
(corrections due to $g_\mu-2$ are included later as the muon self-energy) 
\begin{eqnarray}
\delta_{\rm rel} E_{LS} &=& 
\langle 2P_{1/2}|\delta H_{BP}|2P_{1/2}\rangle -
                 \langle 2S_{1/2}|\delta H_{BP}|2S_{1/2}\rangle \nonumber \\
&=& \frac{\alpha^4\,m_r^3}{48\,m_p^2} = 0.05747\,{\rm meV} 
\label{muh09}               
\end{eqnarray}   
is quite small, as expected. This means that
the perturbative treatment of relativistic effects is 
very appropriate and higher order terms will be even smaller, 
if not negligible.
In deriving (\ref{muh09}), we neglected the hyperfine interaction.
Although we are not interested here in the hyperfine splitting,
this interaction leads to the mixing of $P_{1/2}$ and $P_{3/2}$
states and  shifts these levels. This additional
mixing correction was first considered in \cite{Pachucki:1996:LSmup} 
and amounts to
\begin{equation}
\delta E_{LS}(P_{1/2}^{F=1}) = -0.1446\,{\rm meV},
\end{equation}
and $E_{LS}(P_{3/2}^{F=1})=+0.1446$ meV.
By definition, we do not include this as part of  the Lamb shift,
but consider it as a separate shift of  $P_J^{F=1}$ levels, see Fig. \ref{fig4:muonicH}.

So far we showed that the  $\alpha^2$ relativistic effects
are relatively small. Indeed, the leading effect on the Lamb shift
comes from vacuum polarization effects. Namely, the Coulomb interaction
between the muon (electron) and the proton is modified 
by creation of the electron-positron pairs in the electric 
field. In the language of quantum electrodynamics
the photon propagator includes closed fermion loop corrections,
what results in the modification of the Coulomb interaction
by the so called Uehling potential
\begin{equation}
V_{\rm vp}(r)  =  -\frac{Z\,\alpha}{r}\,
\frac{\alpha}{\pi}\,\int_4^\infty \frac{d(q^2)}{q^2}\,e^{-m_e\,q\,r}\,u(q^2)\,.
\label{muh11}
\end{equation}
where
\begin{equation}
u(q^2) = \frac{1}{3}\sqrt{1-\frac{4}{q^2}}\,\left(1+\frac{2}{q^2}\right)\,.
\end{equation}
The characteristic range of the electron vacuum polarization $V_{\rm vp}(r)$
is about the electron Compton wavelength, which is close 
to the Bohr radius in $\mu$H.
Namely the ratio $\beta = m_e/(m_r\,\alpha) = 0.737386$ is 
not far from one. This means that the parameter $\beta$
must  be kept intact in making expansions in $\alpha$, because the muon nonrelativistic wave function
has a significant overlap with the electronic vacuum polarization potential
$V_{\rm vp}$. As a result, the vacuum polarization correction is 
large for  muonic hydrogen, in contrast to that in the electronic hydrogen. 
Indeed, the leading vacuum polarization correction
\begin{equation}
\delta_{\rm vp} E_{LS} = \langle 2P_{1/2}|V_{\rm vp}|2P_{1/2}\rangle
- \langle 2S_{1/2}|V_{\rm vp}|2S_{1/2}\rangle = 205.0073\,{\rm meV}
\end{equation} 
is the dominating part of the muonic hydrogen Lamb shift \cite{Pachucki:1996:LSmup},
and all other corrections are at least two orders of magnitude smaller.
Note, that the expectation value is taken with nonrelativistic
wave function and the muon-proton mass ratio $\eta$ is included exactly.
At first glance this procedure is less accurate than the expectation value
taken with Dirac wave function,
but we point out that the use
of the reduced mass in the Dirac equation is more than questionable, because
  the leading relativistic recoil correction is not properly accounted for
by using the reduced mass. 
In the present discussion, the vacuum polarization effects are treated
perturbatively by  taking  expectation values instead of
solving numerically the Schr\"odinger equation with this potential.
Higher-order effects obtained by solving the Dirac equation can be included in the 
perturbative approach by including  the higher-order  terms as well.
The slight advantage of perturbative approach is that the vacuum 
polarization potential acquires higher order QED corrections
(two- and three-loops), therefore it is also known perturbatively,
so higher order perturbation and loop corrections are treated on equal
footing. Namely the second order correction due to the Uehling potential,
\begin{equation}
\delta E_{LS}  = 0.1509\,{\rm meV} ,
\end{equation}
is combined with the two-loop vacuum polarization
(one-particle reducible and irreducible two-loop diagrams)
\begin{equation}
\delta E_{LS}  = 1.5081\,{\rm meV}. 
\end{equation}
The last correction is clearly the dominant one.
Although much smaller, the complete $\alpha^3$ correction
due to electronic vacuum polarizations is included also
in present $\mu$H theory.
It was calculated,  first by Kinoshita in \cite{Kinoshita:1999-three-loop},
and later slightly corrected by Karshenboim {\em et al.} 
in Ref.~\cite{Ivanov:2009:three-loop}. The final result is
\begin{equation}
\delta E_{LS}  = 0.0053\,{\rm meV}.
\end{equation}   
The higher order (4-loops or more) vacuum polarization corrections 
are negligible at present. 
There are however additional vacuum polarization
corrections, which are quite interesting.
The electron-positron pairs are distorted by the presence of the
real particles. Namely, these pairs can additionally interact
with the muon or with the proton. This leads to three
box type of diagrams, and the overall effect
as obtained by Karshenboim {\em et al} \cite{Karshenboim:2010:JETP_LBL}
\begin{equation}
\delta E_{LS} = -0.00089(2)\,{\rm meV},
\end{equation}
is almost negligible, as these three diagrams tend to cancel out.
We nevertheless include them in the overall theoretical predictions 
to demonstrate that vacuum-polarization effects
are calculated as completely as possible, and no significant effects
are being neglected. 

There are also vacuum polarization
correction from muon pairs, and this effect is included later together
with the muon self-energy.
Finally the hadronic vacuum polarization, considered in detail
in the context of the muon $g-2$, amounts to \cite{Friar:1999}
\begin{equation}
\delta_{\rm hvp} E_{LS} = 0.0112(4)\,{\rm meV}.
\end{equation} 
   
Up to this point, all the vacuum polarization corrections has been calculated
using the nonrelativistic wave function. The relativistic $O(\alpha^2)$ effects
combined with evp (electron vacuum polarization) 
are included separately as another correction.
Again,  the muon-proton mass ratio $\eta$ is treated exactly,
so the use of Dirac equation is not appropriate. Instead,
we introduce  vacuum 
polarization effects in the  Breit-Pauli Hamiltonian. This is achieved as follows. One notes from 
Eq. (\ref{muh11}) that the evp corrected Coulomb interaction can be represented 
as a Coulomb interaction obtained with the exchange of  massive photons,
integrated with the mass dependent weight function. 
So, one derives the modified Breit-Pauli Hamiltonian $ \delta_{\rm vp}H_{BP} $ obtained using the massive photon, 
neglects the hyperfine interaction and finds that the corresponding 
correction to binding energy is
\begin{equation}
\delta_{\rm vp, rel} E_{LS} = \langle \delta_{\rm vp}H_{BP} \rangle + 
2\,\langle V_{\rm vp}\,\frac{1}{(E-H)'}\,H_{\rm BP}\rangle.
\end{equation}      
The resulting contribution to the Lamb shift
was first calculated in \cite{Pachucki:1996:LSmup}, however with some errors.
The correct result was first obtained by Jentschura in \cite{Jentschura:2011:relrecoil} 
and confirmed by Karshenboim in \cite{Karshenboim:2012:relrecoil} and amounts to
\begin{equation}
\delta_{\rm vp,rel}E_{LS} = 0.01876\,{\rm meV}.
\label{muh20}
\end{equation}
If one used the Dirac equation in the infinite nuclear mass limit, 
the obtained result $0.021$ meV would differ significantly from (\ref{muh20}). 
Nevertheless, this correction is quite 
small, as any relativistic correction for $\mu$H, so higher order
relativistic effects combined with evp can safely be neglected.

What next? In  electronic hydrogen the dominating part
of the Lamb shift is the electron self-energy. In the case
of $\mu$H it is a small but nevertheless important correction.
The formula for the one-loop Lamb shift in $e$H 
is well known \cite{Eides:2006:Book}, the resulting correction in $\mu H$ is
obtained by replacement of the electron mass by the muon mass with the result
\begin{equation}
\delta E = -0.6677\,{\rm meV}.
\end{equation}
Further related corrections arise  from including  both
 muon self-energy and   electron vacuum polarization effects.
The calculation   is not very simple  as it involves
modifying  Bethe logarithms by evp. The complete
calculation was performed by Jentschura \cite{Jentschura:2011:AnnPhys1} 
and the resulting correction is 
 \begin{equation}
\delta E = -0.0025\,{\rm meV}.
\end{equation} 

We pass now, to remaining corrections which have an overlap
with the proton elastic structure  and polarizability effects, and 
not always are treated consistently in the literature.
The pure recoil corrections of order $\alpha^5$,
derived originally by Salpeter, assume that both particles 
have spin $1/2$ and are point-like. They can not be obtained from
the Dirac equation by including  the reduced mass,
as the Dirac energy is even in $\alpha$. Their derivation
requires a full QED treatment and the obtained result~\cite{Eides:2006:Book}
\begin{eqnarray}
E(n,l) = \frac{m_r^3}{m_\mu\,m_p}\frac{(Z\,\alpha)^5}{\pi\,n^3}
& \biggl\{ &\hspace*{-1ex}
\frac{2}{3}\,\delta_{l0}\,\ln\left(\frac{1}{Z\,\alpha}\right)
 - \frac{8}{3}\,\ln k_0(n,l)
-\frac{1}{9}\,\delta_{l0}-\frac{7}{3}\,a_n \nonumber \\ &&\hspace*{-5ex}
-\frac{2}{m_p^2-m_\mu^2}\,\delta_{l0}\,
\left[
m_p^2\,\ln\biggl(\frac{m_\mu}{m_r}\biggr)-
m_\mu^2\,\ln\biggl(\frac{m_p}{m_r}\biggr)\right]\biggr\}\,,
\end{eqnarray}
where
\begin{equation}
a_n = -2\,\left(\ln\Bigl(\frac{2}{n}\Bigr) + \Bigl(1+\frac{1}{2} + \ldots + \frac{1}{n}\Bigr)
+1-\frac{1}{2\,n}\right) \,\delta_{l0} +
\frac{1-\delta_{l0}}{l\,(l+1)\,(2\,l+1)}\,,
\end{equation}
is valid for an arbitrary mass of orbiting particles.
Later, in the next section, we consider the elastic and inelastic
two-photon exchange contribution, and this Salpeter correction
has to be consistently subtracted out from the elastic amplitude.
Coming back to pure recoil effects  in $\mu$H,
the resulting correction is
\begin{eqnarray}
\delta E_{LS} = -0.0450\,{\rm meV}.
\end{eqnarray}

The last correction we consider is due to the proton self-energy
and the related definition of the proton charge radius \cite{Pachucki:1996:LSmup}.
We mention in passing that for an arbitrary spin $I$ nucleus
the charge radius is defined in NRQED by the effective coupling
\begin{equation}
\delta H = 
-e\,\biggl(\frac{\langle r^2_p\rangle}{6} + \frac{\delta_I}{M^2}\biggr)\,
\vec\nabla\cdot\vec E,
\end{equation}
where $\vec E$ is the electric field external to the proton.
For a spinless nucleus $\delta_0 = 0$, for half-spin $\delta_{1/2} = 1/8$,
what corresponds to the so called Darwin-Foldy term. We point out for future 
considerations that for higher spins there is no unique value of $\delta_I$. 
Here, the muon is point-like, so its charge radius by 
definition vanishes, and the charge radius of the proton,
the main issue of this review, is defined by the above equation.
This definition would be equivalent to that obtained from the slope of
$G_E(Q^2)$ form factor at $Q^2=0$, if the electromagnetic form factors
are meaningful at an arbitrarily high precision level. This  is not  necessarily
the case because of QED radiative corrections. 
The proton self energy leads to the modification of
elastic form factors in such a way that they depend on a fictitious
photon mass, or in other words, are no longer well defined.
The only correct definition has to employ complete two-photon 
structure functions, but this will not be analyzed here.
We take the simplest possible point of view and use the formula
for the low energy part of the proton self-energy
\begin{equation}
E(n,l) = \frac{4\,m_r^3\,(Z^2\,\alpha)\,(Z\,\alpha)^4}
{3\,\pi\,n^3\,m_p^2}\left(\delta_{l0}\,
\ln\left(\frac{m_p}{m_r\,(Z\,\alpha)^2}\right)-\ln k_0(n,l)\right)\,.
\end{equation} 
The corresponding correction in $\mu$H is
\begin{equation}
\delta E_{LS} = - 0.0099 \,{\rm meV}\,.
\end{equation}
The high energy part of the Lamb shift is by definition 
included in the proton formfactors,
more precisely in the charge radius and the magnetic moment anomaly.
The problem which remains is how this definition corresponds to
the proton charge radius as obtained from the electron scattering.
We do not aim to analyze it, because possible 
inconsistencies are much smaller than the observed discrepancy
between $\mu$H and the electron scattering determination of $r_p$.  

All corrections presented in this section sum to
\begin{equation}
\Delta E_{LS} = 206.0330\, {\rm meV}.
\label{muH29}
\end{equation} 
There are many further small QED corrections considered in the literature
which are higher order in $\alpha$ or in powers of vacuum polarization.
The main purpose of these further calculations was to find a possible 
missing large correction. However, no significant QED correction has been 
found yet, which would explain the discrepancy, but 
in principle we can not exclude its existence. 

All known corrections have been summarized recently by Antognini {\em et al.} \cite{Antognini:2012:Theory}, and we refer readers 
to that work for more details. Their value (Eq.(\ref{muH30}) below) 
differs only slightly from that in Eq. (\ref{muH29}). 

The calculation of the fine and hyperfine splittings \cite{Martynenko:2008:HFS_Pstates_mup}, 
see Fig. \ref{fig4:muonicH}, except for $E_{\rm HFS}^{2S_{1/2}}$ is much simpler, 
so we present here only the final numerical results \cite{Antognini:2012:Theory}
\begin{eqnarray}
\Delta E_{\rm LS} &=& 206.0336(15) - 5.2275(10)\,r_p^2 + \Delta E_{\rm TPE}\nonumber \\
\Delta E_{\rm FS} &=& 8.3521\,{\rm meV} \nonumber\\
\Delta E_{\rm HFS}^{2S_{1/2}} &=& 22.8089(51)\,{\rm meV},\;\;\mbox{(\rm exp. value)} \nonumber\\
\Delta E_{\rm HFS}^{2P_{1/2}} &=& 7.9644\,{\rm meV}\nonumber \\
\Delta E_{\rm HFS}^{2P_{3/2}} &=& 3.3926 \,{\rm meV} \nonumber\\
\Delta                &=& 0.1446\,{\rm meV}
\label{muH30}
\end{eqnarray}
where $\Delta E_{\rm TPE} = 0.0351(20)$ meV (see Table 1) is a proton structure dependent two-photon
exchange contribution, which is considered in the next section.
The theoretical value for $2S_{1/2}$ hyperfine splitting is uncertain
due to not well known proton-structure effects in magnetic interactions,
so the above value is the experimental one, taken from \cite{Antognini:2013:Science_mup2}.

%
%
\subsection{Proton Polarizability Effect in Muonic Hydrogen}
\label{sec:polarizab}

\begin{figure}[b]
\begin{center}
\includegraphics[width = 0.5\columnwidth]{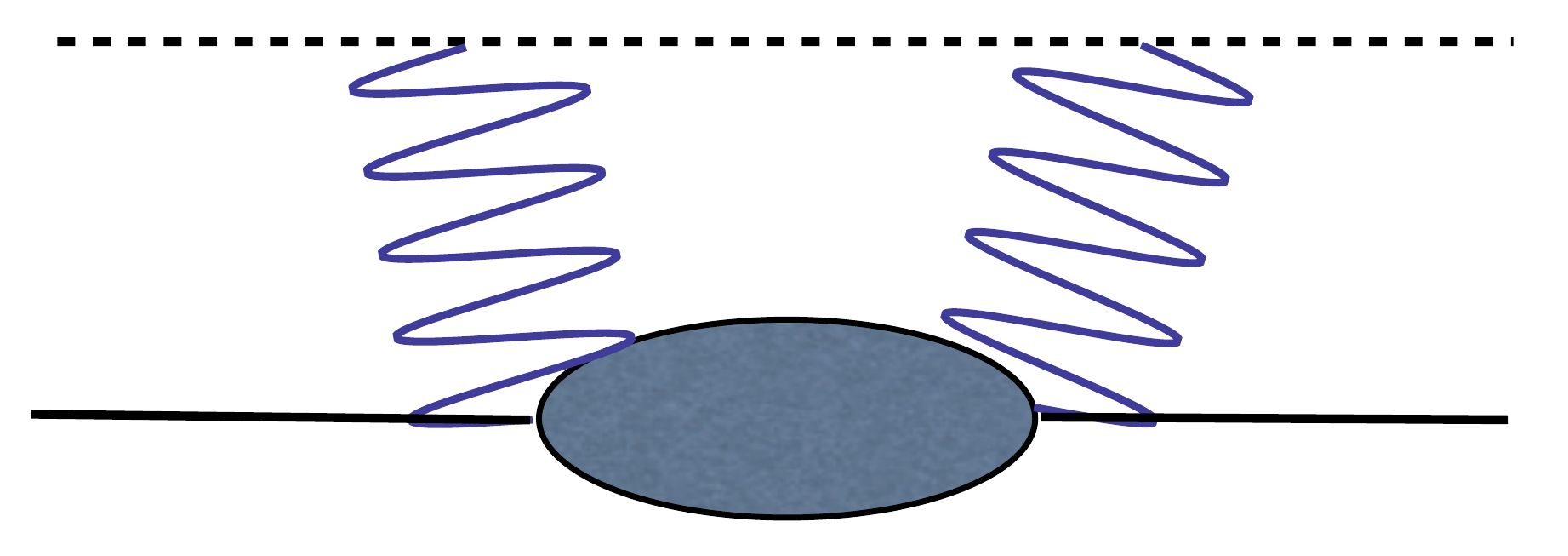}
\caption{The box diagram for the $\mathcal O(\alpha^5m^4)$ corrections. 
  The graph in which the photons cross is also included in the calculation.
  The blob represents all possible excitations of the proton, the wiggly lines
  represent the exchanged photons.
  The solid line represents the proton and the dashed line represents the muon.
}
\label{fig5:lambbox}
\end{center}
\end{figure}
 
The previous subsection is concerned with the QED calculations needed to
determine the proton radius from the muonic energy level splittings between
$2P$ and $2S$ states. There is one particular term, occurring at the
intersection between QED and strong interaction effects which is deserving of
special attention. This is the proton polarizability contribution that enters
in the two-photon exchange term, see Fig.~\ref{fig5:lambbox}.  
The computed effect of this term is proportional to the lepton mass to the
fourth power, and so is capable of being relevant for muonic atoms, but
irrelevant for electronic atoms.

The proton can be excited by the absorption of a photon and then de-excited by
the emission of a photon. Thus the two-photon exchange term depends on the
forward virtual photon-proton (Compton) scattering amplitude $T^{\mu\nu}(\nu,
q^2)$ where $q^2$ is the square of the four momentum, $q^\mu$ of the virtual
photon and $\nu$ is its time component. The contribution of the graph of
Fig.~\ref{fig5:lambbox} is given by the
expression~\cite{Pachucki:1999:ProtonMup}:
\bea
E=-{e^4\over 2}\phi(0)^2\int {d^4q\over (2\pi)^4i}{1\over q^4}[T^{\mu\nu}-t^{\mu\nu}(M)]t_{\mu\nu}(m),
\label{twog}
\eea
where $t^{\mu\nu}$ is the Compton amplitude for a pointlike fermion of mass
$M$ (proton) or $m$ (muon), $ \phi^2(0)={\alpha^3 m_r^3\over 8\pi}$ for the 2S
state with $m_r$ as the lepton reduced mass.  The role of the term
$t^{\mu\nu}(M)$ in \eq{twog} is to subtract the term recoil correction
discussed in the previous section. Using \eq{twog} still requires the
subtraction of the infrared divergent term related to the QED correction to
the proton charge radius discussed in the previous section.

      The quantity  $T^{\mu\nu}(\nu, q^2)$ is decomposed  into  a combination of  two  independent terms,  $T_{1,2}(\nu,q^2)$, allowed by symmetries.  The imaginary parts of the functions 
$T_{1,2}(\nu,q^2)$ are  related to structure functions  $F_{1,2}$ measured in electron- or muon-proton scattering, so that
$T_{1,2}$ can be expressed in terms of $F_{1.2}$ through dispersion relations. 
However,
 $F_1(\nu,Q^2)$  falls off too slowly for large values of $\nu$ for the dispersion relation to converge. Hence, one makes  a 
 subtraction at $\nu=0$, requiring that a new function of $Q^2$, $\overline T_1(0,Q^2),$  be introduced. The contributions to the Lamb shift are broken into three terms~\cite{CarlsonVanderhaeghen:2011}:
  $E=\Delta E^{el}+\Delta E^{inel}+\Delta E^{subt}$, where $el$ refers to  lepton-proton intermediate states, $inel$ refers to  intermediate excitations of the proton, and $subt$ refers to the subtraction term. This separation into three terms is somewhat arbitrary, as emphasized by Ref.~\cite{HillPaz:2011:TwoPhotons}. But so far no authors have  found
 that differences between different possible separations are relevant for resolving the proton radius puzzle. 
  
  The subtraction term has typically been handled by 
  using a power series expansion around $Q^2=0$, and then using effective field theory 
 to determine the coefficients of the series. In practice, people~\cite{CarlsonVanderhaeghen:2011,Pachucki:1999:ProtonMup,Martynenko:2006:Pol_mup_H,BirseMcGovern:2012} have used 
 \be
\overline T_1(0,Q^2) = \frac{\beta_M}{ \alpha} Q^2 F_{\rm loop}(Q^2),	\,.\label{six}
\ee
where $F_{\rm loop}(Q^2)$ is a function that falls with $Q^2$ sufficiently rapidly so that the necessary integral converges.
The calculations of three groups are summarized in 
   Table~\ref{table:twog}.  
\begin{table}[htdp]
\caption{Numerical results for the $\mathcal O(\alpha^5)m^4$ proton structure corrections to the Lamb shift in muonic hydrogen.  Energies are in $\mu$eV.}
\begin{center}
\begin{tabular}{lccc}
($\mu$eV)	&	Ref~\cite{CarlsonVanderhaeghen:2011} &  Ref.~\cite{Pachucki:1996:LSmup,Pachucki:1999:ProtonMup}			& Ref.~\cite{Martynenko:2006:Pol_mup_H} \\
			\hline
$\Delta E^{subt}$  & $\quad\ \, 5.3 \pm 1.9$ & $\quad\ \, 1.8$ & $\quad\ \, 2.3$ \\
$\Delta E^{inel}$  & $-12.7 \pm 0.5$  & $-13.9$ & $-16.1$ \\
$\Delta E^{el}$  & $-29.5 \pm 1.3$ & $-23.0$ & $-23.0$ \\
\hline
$\Delta E$  & $-36.9 \pm 2.4$ & $-35.1$ & $-36.8$ \\
\end{tabular}
\end{center}
\label{table:twog}
\end{table}
The theoretical uncertainty as determined from the   three values of $\Delta E$ shown in the Table would seem to be very small. However, there is a hint of trouble in the large relative differences between the values for $\Delta E^{subt}$. Birse \& McGovern~\cite{BirseMcGovern:2012} reevaluated the subtraction term (using the separation of terms employed by
Ref.~\cite{CarlsonVanderhaeghen:2011})
and computing  one order higher  in the power series for $Q^2$, to find $\Delta E^{subt}=4.2 \pm1.0 \mu$eV.  Some authors~\cite{HillPaz:2011:TwoPhotons,Miller:2011yw,Miller:2012:quasi-elastic,Miller:2012ne} have argued that the uncertainty in this  term is substantial, even though the
typical values are  only a few $\mu$eV, very small compared to the 310 $\mu$ eV needed 
to account for the proton radius puzzle. 
  
The trouble is that the integral that determines $\Delta
E^{subt}$~\cite{Miller:2012ne} would diverge logarithmically without the
introduction of the form factor $F_{\rm loop}(Q^2).$ In principle, one needs
to evaluate an infinite set of terms in the power series describing $\overline
T_1(0,Q^2)$ and hope that the result would lead to a convergent integral.
Miller~\cite{Miller:2012ne} used the form \be F_{\rm loop}(Q^2)={Q^4\over
  M_\gamma^4}{ 1\over (1+ a Q^2)^3 }.\label{mine}\ee The use of this form
allows one to state the expression for the energy shift in closed form, which
is well approximated by
\bea
\Delta E^{subt} \approx  {3\over2}m{\alpha \phi^2(0)\; } {\beta_M\over M_\gamma^4a^2}.
\label{eq:ppol:res}
\eea 
Using, $M_\gamma=0.5$ GeV and $a=0.0078 /(4m_\mu^2) =$0.177 GeV$^{-2}$ led to
value of the Lamb shift of 0.31 meV.  A relatively large value of $1/a$ must
appear to obtain the necessary contribution, but no fundamental principles
have been violated.

Thus the above discussion shows that procedures used to estimate the size of
the subtraction term suffer from significant uncertainties. This arises
because the chiral EFT is being applied to the virtual-photon nucleon
scattering amplitude, and as a result the energy shift depends on an integral
that would be logarithmically divergent if the function $F_{\rm loop}$ were
replaced by unity.
Another way to proceed would be us to use an effective field theory (EFT) for
the lepton-proton interaction.

In EFT logarithmic divergences identified through dimensional regularization
are renormalized away by including a lepton-proton contact interaction in the
Lagrangian.
This was done~\cite{Miller:2012ne} by using standard dimensional
regularization (DR) techniques by evaluating the scattering amplitude of
Fig.~\ref{fig5:lambbox}.
The term of interest is obtained by including only $\overline T_1(0,Q^2)$ of
\eq{six} with $F_{\rm loop}=1$.  The result is :
\bea
{\cal M}_2^{DR}({\rm loop}) ={3\over2}i\; \alpha^2 m{ \beta_M\over \alpha}\big[{2\over \epsilon}+\log {{\mu^2\over m^2}}+{5\over 6}-\gamma_E+\log 4\pi\big]\overline u_f  u_i  \overline  U_f U_i ,
\label{eq:ppol:res1}
\eea
where lower case spinors represent leptons of mass $m$, and upper case proton
of mass $M$, $\gamma_E$ is Euler's constant, 0.577216$\cdots\;$, $\mu$ is the
parameter introduced in using dimensional regularization and one works in a
space of dimension $d=4-\epsilon$.

The result \eq{eq:ppol:res1} corresponds to an infinite contribution to the
Lamb shift in the limit that $\epsilon$ goes to zero. In EFT one removes the
divergent piece by adding a lepton-proton contact interaction to the
Lagrangian that removes the divergence, replacing it by an unknown finite
part.
The finite part is obtained by fitting to a relevant piece of data.  Here the
only relevant data is the 0.31 meV needed to account for the proton radius
puzzle. When this is done the result is
\bea
{\cal M}_2^{DR} =i\; 3.95 \;\alpha^2 m {4\pi\over \Lambda_\chi^3 }\overline u_f  u_i  \overline  U_f U_i .
\label{mdr1}
\eea
where the results are expressed in terms of the chiral symmetry breaking
scale$\Lambda_\chi \equiv 4\pi f_\pi$, ($f_\pi$ is the pion decay
constant). The coefficient 3.95 is of natural size.
Thus standard EFT techniques result in an effective lepton-proton interaction
of natural size that is proportional to the lepton mass.

The present results, \eq{eq:ppol:res} and \eq{mdr1} represent an assumption
that there is a lepton-proton interaction of standard-model origin, caused by
the high-momentum behavior of the virtual scattering amplitude, that is
sufficiently large to account for the proton radius puzzle.
This assumption needs further testing.  Fortunately, our hypothesis can be
tested in an upcoming low-energy $\mu^\pm p, e^\pm p$ scattering experiment
that is discussed in Sect.~5.2.

%
%
\subsection{Theory of hydrogen energy levels}

One of the possible explanations for the proton charge radius puzzle is that it is caused by 
a mistake or a missing correction in the hydrogen Lamb shift theory.
We will argue  that this is the least probable solution.
Why? It is because corrections to the hydrogen Lamb shift have
been calculated by many groups using different methods, 
with only few exceptions which are described below in detail.
For a recent review of Hydrogen theory in the context
of determination of the proton charge radius see Ref.~\cite{Mohr:2012:CODATA10}. 

All the corrections to the Lamb shift are classified in powers of
$\alpha$, $Z\,\alpha$ and the mass ratio $m_e/m_p$,
where the power of $\alpha$ counts the number of loops.  
The additional 
proton structure corrections, elastic and inelastic two-photon
contributions are of the order of tens of Hz for 1S state 
\cite{Mohr:2012:CODATA10}, 
while the discrepancy in the proton charge radius corresponds 
to about 100 kHz. So, any proton structure effect beyond the 
the finite size correction is negligible. 

The leading contributions to the hydrogen Lamb shift, the one-loop
electron self-energy and the vacuum polarization have been calculated
by an expansion in $Z\,\alpha$ \cite{Jentschura:2005:NRQED} or 
directly numerically \cite{Jentschura:2004:SELF},
using the known form of the relativistic electron propagator
in the Coulomb field. Both methods are in agreement, but the 
numerical one is presently more accurate, about a few Hz for the ground state,
so there is  no room for any mistake of order of 100 kHz.

The two-loop contribution has also been calculated numerically
\cite{Yerokhin:2010:TWLP} and analytically 
\cite{Pachucki:2003:TWLP, Jentschura:2005:NRQED}, 
but the results differ slightly. Numerical calculations
have been performed only for the nuclear charge $Z\geq10$
and the result for $Z=1$ was obtained by extrapolation.
The analytic calculation was numerically accurate, but some 
higher order terms were omitted, because they  are  too
difficult to  evaluate. The final result is taken as
an arithmetic mean with an uncertainty that matches both values.
This uncertainty corresponds to about 2 kHz for 1S state, much too small
to explain the discrepancy.

The three loop contribution in the leading order of $Z\,\alpha$
has been calculated only by one group \cite{Melnikov:2000:THREE}.
This correction can be
represented in terms of three-loop electromagnetic form factors of the 
free electron. Although technically very advanced,
the conceptually simple methods of the evaluation of
Feynman diagrams can be applied, so we do not expect a mistake here.
The binding corrections, next order in $Z\,\alpha$ are known
only partially \cite{Eides:2007:THREE},
but can be roughly estimated to be about $\sim$ 1 kHz.

The pure recoil effects, two-body diagrams without self-energy
and vacuum polarization, although small, are  not very simple to 
evaluate because the  full QED formalism must be used. 
The derivation of the closed formula 
for the pure recoil correction was achieved independently
by two groups \cite{Shabaev:1985:REC, Pachucki:1995:REC}. 
The numerical evaluation was performed by analytic expansion in $Z\,\alpha$
\cite{Pachucki:1995:REC,Eides:1997:REC} and directly numerically \cite{Shabaev:1998:REC}.
Both are in agreement. 

The radiative recoil corrections have been calculated only by $Z\,\alpha$ 
expansion, but independently by two groups
\cite{Pachucki:1995:RREC,Eides:1995:REC}.
The overall result is rather small, some 
logarithmic higher order terms are also known \cite{Mohr:2012:CODATA10},
and remaining higher corrections are irrelevant at the order
of discrepancy.

In conclusion there is no room for any new correction,
or any mistake in the present calculations that would
shift 1S Lamb shift by about 100 kHz.

%
%
 
\subsection{Physics Beyond the Standard Model}
The possibility that the proton radius puzzle might be caused by a difference between the $\mu p$ and $ep$ interactions is very exciting because such an effect is not part of  the Standard Model (SM). Awareness of the possibility of such a failure  of what is known as electron-muon (or more generally lepton)  universality was re-awakened by the muon $g-2$ 
experiment~\cite{Hertzog:2004gc,Bennett:2006fi,Jegerlehner:2009ry}. The experimental value of the muon anomalous magnetic moment $a_\mu$ exceeds the  SM expectation by more than three standard deviations: $\Delta a_\mu\equiv a_\mu^{\rm exp}-a^{\rm th}_\mu=(29\pm9)\times 10^{-10}$. This difference could be caused by a new interaction.
In addition, new gauge forces mediated by particles in the MeV-GeV scale could be dark matter candidates~\cite{Fayet:2004bw,Finkbeiner:2007kk,Pospelov:2007mp}.
 Obtaining  a theory or  model that could account for the proton radius puzzle, the muon anomalous magnetic moment and   predict other new  verified phenomena would be a tremendous breakthrough.
However, it is not  easy to invent new interactions that differentiate between electrons and muons because   the  principle of lepton universality is very well tested. 

The most relevant constraints of lepton universality have been nicely summarized in Refs.~\cite{Barger:2011,Barger:2012}.
The basic ideas are to use the non-observation of new particles that couple only to muons and  to search for differences between muonic and electronic decays of various particles.
Constraints are obtained from 
  the decay of the $\Upsilon$ resonances~\cite{Aubert:2009cp,Wilczek:1977zn,delAmoSanchez:2010bt};
 neutron interactions with nuclei~\cite{Barbieri:1975xy,Schmiedmayer:1991zz,Leeb:1992qf}; the anomalous magnetic moment of the muon~\cite{Jegerlehner:2009ry}; x-ray transitions in $^{24}$Mg and $^{28}$Si atoms; $J/\Psi$ decay~\cite{Insler:2010jw}; neutral pion  decay~\cite{Altegoer:1998qta}; and eta decay~\cite{Ablikim:2006eg}.

Barger {\it et al.}~\cite{Barger:2011} postulate the existence of new interactions mediated by the  exchange of a boson between   muons and nucleons.
To explain the proton radius puzzle, these exchanges  must give rise to an attractive interaction that accounts for 0.31 meV. One can include the exchange of scalar, vector and tensor bosons. Axial vector exchange gives a spin-spin interaction that affects the hyperfine splitting, but not the Lamb shift, and is not considered further  in Ref.~\cite{Barger:2011} .

Given the non-relativistic motion of the muon and the nucleon, the exchange of boson between the muon and nucleon leads to a Yukawa interaction                                            :
\bea \Delta V(r)=-\alpha_\chi{e^{-m_\chi r}\over r},\eea 
where $m_\chi$ is the mass of the postulated boson and $\alpha_\chi$ is the product of lepton- and nucleon-boson coupling constants divided by $4\pi$. Assuming sufficiently weak couplings, first order perturbation theory can be used to obtain the contribution to the muon Lamb shift~\cite{BorieRinker:1982:muAtoms} as
\bea \delta(\Delta E)=\alpha_\chi m_\chi {{m_\chi\over \alpha m_r}\over 2(1+{m_\chi\over\alpha m_r})^2},\label{eq:BSM:res}\eea
where the muon-nucleon reduced mass is $m_r$. 

Barger {\it et al.} conclude that new spin 0,1 and 2 particles that mediate flavor-conserving non universal spin-independent interactions are excluded. This finding depends on assuming that the new particles couple only to the muon, and that the coupling to mesons is the same as to the nucleon.

Brax \& Burrage~\cite{Brax:2011} note that light scalar mesons coupled to matter appear in many fundamental contexts, 
such as in the inflationary scalar field and in attempts to  unify gravity with the standard model. However, these interactions are strongly constrained by experimental searches for fifth forces, and violations of the equivalence principle. Non-linear dynamics~\cite{Khoury:2003rn,Khoury:2003aq,Nicolis:2008in} can be introduced to avoid these constraints. Brax \& Burrage  use  the assumption that matter couples universally to the scalar field to  obtain a new stringent bound on the coupling of  scalar
fields to matter from the precise measurements of  the difference between the energies of the 1s and  2s states in hydrogen, while the coupling to photons is constrained by the Lamb shift. The net result of these considerations is that the contribution of a scalar field to the
discrepancy between proton radii measured in electronic and muonic atoms is negligible. 

Tucker-Smith \& Yavin~\cite{Tucker-Smith:2011} go through   a  chain of reasoning similar to that of  Barger {\it et al.}~\cite{Barger:2011}, and find that
a new force carrier with mass of order MeV that couples to protons and muons can account for the proton radius puzzle as well as  the difference between QED and measurements for $g-2$ of the muon. They acknowledge that many constraints exist, but state that none of the existing constraints exclude such a force in a model-independent manner. The difference in conclusion with Barger {\it et al.}  arises because Tucker-Smith \& Yavin do not assume universal couplings to nucleons and mesons. Tucker-Smith \& Yavin state that their postulated force can be tested in measurements of muonic-deuterium and helium.

Motivated by the interest in in the possibility of new gauge forces the mediate interactions between dark matter and standard model parameters, Batell {\it et al.}~\cite{Batell:2011:PV_muonic_forces} use a model in which a new $U(1)$ gauge boson
(heavy photon) is kinetically mixed with hypercharge. New scalar and vector force carriers interact with right-handed neutrinos. Current experiments at JLab, Mainz, KLOE and BaBar are searching for such particles but haven't found any. This limits the value of the mass of the heavy  photon
to be about 30 MeV.  If the strength of this particular new force is large enough to account for the proton radius puzzle, parity violating asymmetries in muon-nucleon or muon-nucleus scattering would be enhanced by several orders of magnitude and would be of order $10^{-4}$. Ref.~\cite{Barger:2012}  use the nonobservation of missing mass events in the lepton kaon decay
$K\to\mu\chi$ to find strong constraints on parity-violating gauge interactions of a right handed muon provided $\chi$ decays invisibly or does not decay inside the detector.

Carlson \& Rislow~\cite{CarlsonRislow:2012:NewPhysisc} consider two models, one containing  scalar and pseudo scalar particles and the other vector and axial vector particles. The basic idea is that the scalar or vector particles account for the Lamb shift while the pseudoscalar and axial vector particles do not contribute.  Large effects in the muon $g-2$ are prevented by  cancellations   between the effects of scalar (vector) and pseudoscalar (axial) mesons. Constraints on the masses of the new particles are obtained from the nonobservation of such particles in kaon decays. Exchanges of  pseudoscalar and axial vector particles  contribute to  hyperfine splittings, and the resulting constraints are not considered in Ref.~\cite{CarlsonRislow:2012:NewPhysisc}.

It is also worthwhile to discuss the possible influence of new forces on the PSI muon-proton scattering experiment (Sect.~\ref{sec:mu_p_scatt}). If the new particles are light, they act like a very small modification of the Coulomb force and can not be detected. If the new particles are sufficiently  heavy, the force between the muon and the proton is well-approximated by a contact interaction (as in the model~\cite{Miller:2012ne} discussed in Sect. ~3e) and  its effects are detectable as an interference between a short-ranged and long-ranged Coulomb interaction.

To summarize, there are many constraints on new forces. Some of these constraints are derived from model-dependent considerations,
and  the new force possibility remains viable for a subspace of all of the relevant parameters. The ultimate disposition of this possibility will depend on either directly detecting the 
postulated particles or directly verifying an independent prediction of one of these models.  

\section{New projects}
There is an interesting set of new experiments that could shed light on the proton radius puzzle.

%
%
\subsection{New electron scattering experiments}

Improving the precision of the radius determination from
electron scattering experiments
requires that cross sections be measured with higher precision 
and/or to smaller $Q^2$.
All higher precision measurements that have been done to date
have used magnetic spectrometers with small or moderate
solid angle acceptance, and it is hard to see how a
significant improvement can be made  in the precision 
of these measurements.
As $Q^2$ $\approx$ $4EE^{\prime}\sin^2(\theta/2)$,
going to smaller $Q^2$ requires smaller scattering angles
and/or beam energies.
But the magnetic spectrometers have limited angular ranges, 
and the experiments already run have used the lowest beam 
energies available at modern accelerators.
Thus, an improved electron scattering experiment requires
a different technique.

The only experiment currently approved to improve the electron 
scattering radius determination is Jefferson Lab E12-11-106
\cite{gasparian2011}, which aims to extend the $Q^2$ range of 
electron scattering experiments from the 0.0038 GeV$^2$ of the 
Mainz experiment down to about 10$^{-4}$ GeV$^2$.
The experiment has several novel features that combine to allow
a significant radius determination from a measurement only at low $Q^2$.
Low current, 1 and 2 GeV  beams in Hall B impinge upon an open-ended hydrogen 
gas cell, as has been used in storage rings previously, to eliminate 
the need to subtract target end cap backgrounds.
The spectrometer, the PRIMEX HYCAL (PbWO$_4$) calorimeter,
will cover the entire angle range of $\approx$10 mrad - 70 mrad in one
setting, to cover the $Q^2$ range $10^{-4}$ GeV$^2$ $\to$ $10^{-2}$ GeV$^2$.
Moller scattering of the beam from atomic electrons will
provide the data used to determine the detector solid angles, offsets,
and efficiencies at the high level needed.
The lower energy of the Moller scattered electrons allows the two
reactions to be distinguished by the calorimeter.

The difference in the cross sections arising from the difference
in the measured radii is small, requiring high precision cross
section measurements.
Approximating $G_E = 1 - Q^2 r_p^2 / 6$, 
radii of 0.842 fm vs.\ 0.875 fm lead to differences in the
cross sections of 0.005\%, 0.05\%, 0.5\% and 1.0\% for
$Q^2$ of $10^{-4}$ GeV$^2$,  $10^{-3}$ GeV$^2$, 0.01 GeV$^2$
and 0.02 GeV$^2$, respectively, and differences half that large
for the form factor.
Existing electron experiments each are about 5$\sigma$ different
from the muonic hydrogen result.
If the radius is to be determined to give a similar quality result,
then relative cross sections need to be determined at the 0.2\% level.
The large cross sections and event rates for low $Q^2$ scattering 
are large, making statistical uncertainties at the 0.1\% quite feasible.

Systematic uncertainties must also be controlled at a level that
would be very difficult without a well understood calibration
reaction.
Reaching low $Q^2$ requires small scattering angles and/or
beam energies.
With a 1 GeV beam energy, reaching $10^{-3}$ ($10^{-4}$) GeV$^2$ 
requires a scattering angle of 32 (10) mrad.
But the Mott cross section diverges at small angles due to the
$1/\sin^4(\theta/2)$ factor, leading to a
shift in the cross section of 0.01\%/$\mu$rad (0.03\%/$\mu$rad)
at $10^{-3}$ ($10^{-4}$) GeV$^2$.
A 10 $\mu$rad knowledge of the scattering angle is then needed
to limit shifts the relative cross sections by 0.2\%.
In contrast, typical high precision electron scattering experiments 
with magnetic spectrometers determine absolute scattering angles 
at about the 1 mrad level.

Jefferson Lab E12-11-106 is approved and tentatively scheduled
to run in 2014-2015.
If various technical challenges are met, it has an excellent chance 
of resolving all experimental questions
related to data normalizations and fits used to extract the proton
radius from electron scattering experiments, either
confirming the puzzle, or resolving it as being due to incorrect
radius extractions from the electron (scattering) data.



%
%
\subsection{Elastic muon-proton scattering}
\label{sec:mu_p_scatt}

The missing measurement in determining the proton
radius with muons and electrons is muon scattering,
a measurement that the MUon proton Scattering Experiment
(MUSE) collaboration proposes at PSI \cite{MUSE2012}.
Muon scattering is not an easy measurement, as
muons are produced as secondary beams from
the decays of $\pi$'s produced from a primary proton beam.
As a result, muon beams have low flux and large
emittance, compared to electron beams.
Unless efficient particle separators are available, 
the muon beam will also include background pions and electrons.

The proposed MUSE experiment is based at the PSI $\pi$M1 beam line. 
The beam line provides of order MHz muon fluxes 
for momenta from about 100 $\to$ 250 MeV/$c$.
To handle the multiple particle species in the beam,
MUSE uses Scintillating Fiber arrays to determine RF time
and only operates at momenta where the muon RF time is about 4 or more
ns different from electron and pion RF times.
A custom FPGA system will allow the RF time to be determined in 
hardware so that the information is available to the trigger.
To avoid issues arising from accidental coincidences,
MUSE intends to limit the channel acceptance to keep the
total beam flux at $\approx$5 MHz.
The large emittance of the beam requires the use
of GEM chambers to track individual beam particles
into the target, so that the scattering angle is sufficiently well known.
Test measurements have largely verified that the
beam properties are sufficient for the experiment.

The low rates also require a large acceptance spectrometer.
MUSE is planning a nonmagnetic spectrometer,
consisting of wire chambers and fast scintillators
for triggering and particle identification.
The main reason for this is that experiments with
large acceptance detectors and magnetic fields
do not measure precise cross sections.

MUSE intends to measure with both $\mu^+$ and
$\mu^-$ beams, so that two-photon effects can be directly
determined from the data.
MUSE also intends to measure both muon and electron
scattering at the same time, so that the experiment
can compare them.

The collaboration has studied singles rates, background
triggers, and systematic uncertainties to estimate
how well the cross sections, form factors, and radii
from the various measurements can be compared.
A number of difficulties arise that are not
present in electron scattering experiments.
The $\pi$ induced backgrounds are efficiently removed
at the trigger level through the measurements of beam 
RF time, which is only possible due to the low beam flux.
The beam muons decay, which is mainly a problem as
the decay electrons cannot be distinguished from the
muon scattering at the trigger level, and are not as easily
distinguished with a non-magnetic spectrometer.
Although the muon decay fraction is small, it is much larger
than the scattering fraction.
The muon radiative tail is very small, but electrons
have a large radiative tail which must be handled.
The beam emittance is large, requiring measurement by detectors
in the beam, which further increases the emittance.
Many of the systematic uncertainties are similar
to electron scattering experiments, but the low energy
muon beam has reduced sensitivity to angle offsets,
radiative corrections, and hadronic backgrounds
compared to a GeV electron beam.
Preliminary estimates are that the radius can be
extracted from muon scattering about as well
as in the Mainz experiment, in part because the
experiment can reach $Q^2$ $\approx$ 0.002 GeV$^2$, 
about half the lower limit of the Mainz measurement.

The MUSE experiment requires significant new
funding ($\approx$2 M\$) to be carried out. 
It might be able to run as soon as 2016.

%
%
\subsection{Spectroscopy of electronic atoms and ions}

As detailed in Sect.~\ref{sec:H_exp}, the value of \rp{} can be determined
from the 1S-2S transition in H~\cite{Parthey:2011:PRL_H1S2S} using precisely
calculated QED theory only if a sufficiently precise value of the Rydberg
constant \Ryd{} is supplied. At present, \Ryd{} originates from several
measurements of transition frequencies in H and D.

In fact, the correlation coefficient between \Ryd{} and \rp{} is 0.984 in the
2010 CODATA adjustment~\cite{Mohr:2012:CODATA10}. Hence, any new determination
of the Rydberg constant at a level of a few parts in $10^{12}$ would help to
shed new light on the proton radius discrepancy between electronic and muonic
hydrogen.

Several such projects are underway. Flowers {\it et al.} at the British NPL
are measuring the 2S-6S/D 2-photon transitions in atomic
H~\cite{Flowers:NPL:2007}.  Beyer {\it et al.} at MPQ, Garching~\cite{Beyer:2013:AdP_2S4P}, aim at an
improved value of \Ryd{} from 1-photon transitions in H, namely the 2S-4P and
higher P-states. This project utilizes for the first time a cryogenic beam of
H atoms from a nozzle at 6\,K, optically excited to the 2S state using the
1S-2S laser system~\cite{Parthey:2011:PRL_H1S2S}. This can eliminate possible
systematic effects like dc-Stark shifts from patch charges caused by the
electron bombardment traditionally used to create H(2S) atoms and align the H
atom beam onto the laser beams. The cold atom beam can help to reduce
systematic effects caused by the large velocity of traditional ``hot'' beams,
but the 1-photon transition is particularly sensitive to 1st order Doppler
shifts.
The 1S-3S transition in H has recently been measured by Nez {\it et al.} at
LKB, Paris~\cite{Arnoult:2010:1S3S}. It is the second best measured transition
frequency in H, but as can be seen in Fig.~\ref{fig2:Rp_from_H} it does not
provide the most precise \rp{} value when compared with 1S-2S following
Eq.~(\ref{eq:E_simple}). This is because both the 1S-2S and the 1S-3S
transition depend on the large 1S Lamb shift $L_{1S}$, and the ``lever arm''
on \Ryd{} is hence reduced. Nevertheless, this measurement is very valuable
because it serves as a cross-check for the 1S-2S transition in
H~\cite{Parthey:2011:PRL_H1S2S} which has for a long time been measured
by only the MPQ group. The Paris group is improving their 1S-3S setup, in
particular the laser system, and aims at a significant improvement of their
accuracy~\cite{Arnoult:2010:1S3S}.
The 1S-3S transition in H is also being measured at MPQ, Garching, by Peters
and coworkers~\cite{Peters:2009:comb205nm}. To overcome difficulties in the
generation of the 205\,nm light required to drive the 1S-3S transition in H
they use two counter-propagating picosecond frequency combs.

Laser spectroscopy of He atoms and H-like He$^+$ ions are pursued by 
Udem and colleagues at MPQ~\cite{Herrmann:2009:He1S2S} and by 
Eikema and colleagues~\cite{Kandula:2011:XUV_He} and
Vassen and colleagues~\cite{Rooij:2011:HeSpectroscopy},
both in Amsterdam.
Ultimately, this will allow a Rydberg constant determination using He atoms or
ions. A precise value of the He nuclear charge radius is available from
electron scattering experiments~\cite{Sick:2008:rad_4He}. Soon, the Lamb shift
in muonic helium ions\cite{Antognini:2011:Conf:PSAS2010} will improve this
accuracy further (see Sect.~\ref{sec:exotic_atoms}).

An alternative route towards a new determination of \Ryd{} is pursued by Tan
and coworkers at NIST~\cite{Tan:NIST:2011}. Here, H-like highly charged ions
(e.g.\ Ne$^{9+}$) are created in circular large-$\ell$-states at high principal
quantum number $n$ and stored in a Penning trap. In large-$\ell$-states
the effect of the finite nuclear size on the energy levels is negligible, and
measuring these highly charged ions at high $n$ states ensures that the
transition frequencies can be determined using lasers in the optical region.

Hessels {\it et al.}~\cite{Vutha:2012:H2S2P} at York University, Toronto, 
aim at a new measurement of
the ``classical'' 2S-2P Lamb shift in H. This radio-frequency transition was 
last measured in 1994~\cite{Hagley:1994:FShyd}. The new
direct measurement of the Lamb shift will yield a value of \rp{} that does not
require the knowledge of \Ryd, so it is an independent cross-check of the
optical 2S-$n\ell$ transitions in H and D that determine \Ryd{} today 
(see Sect.~\ref{sec:H_exp}).

%
%

%
%
\subsection{Spectroscopy of exotic atoms}
\label{sec:exotic_atoms}

Spectroscopy of exotic atoms will also contribute to the solution of the
proton radius puzzle. Precision laser spectroscopy of purely leptonic systems
like positronium (Ps~$\equiv e^+e^-$)~\cite{Crivelli:2011:Conf:PSAS2010,Cassidy:2012:Ps_HFS} or muonium
(Mu~$\equiv \mu^+e^-$)~\cite{Antognini:2012:MuoniumEmission} can test
bound-state QED free from finite nuclear size effects. Improved spectroscopy
of the 1S-2S transitions and the ground state hyperfine splittings are on the
way or being prepared. Ps and Mu test QED in the electronic and muonic
sectors, respectively.

Laser spectroscopy of muonic deuterium atoms will soon shed new light on the
radius puzzle~\cite{Pohl:2011:Conf:ICAP2010}. A new charge radius of the
deuteron can be compared with the one obtained from the isotope shift of the
1S-2S transition of electronic H and D~\cite{Parthey:2010:PRL_IsoShift}.  The
muonic deuterium measurement may also give new insights into the question of
nuclear polarizability (see Sect.~\ref{sec:polarizab}).

The PSI project R-10.01 is going to measure the Lamb shift in muonic helium
ions $\mu^3$He and $\mu^4$He~\cite{Antognini:2011:Conf:PSAS2010}. The
measurement will give a tenfold improved value of the charge radii of the
stable helium isotopes, compared to the value from electron
scattering~\cite{Sick:2008:rad_4He}. Together with the new measurements of
transitions in electronic He atoms or
ions~\cite{Herrmann:2009:He1S2S,Kandula:2011:XUV_He,Rooij:2011:HeSpectroscopy}
the muonic
helium Lamb shift will provide a cross-check of the proton radius puzzle at
Z=2.

A future possibility may be the measurement of transitions in muonic Li, Be
and B ions~\cite{Drake:1985:muLi_Be_B}. Accurate charge radius differences of
these nuclei are known from electronic isotope shift
measurements~\cite{Nortershauser:2011:Li_iso_shift,Nortershauser:2009:Rad_7_9_10Be,Krieger:2012:Rad_12Be}.

\section{Conclusions}
%
%

The proton radius puzzle is real. A physical parameter should not depend on the method of extraction.  Yet here we have two highly precise methods based on
electronic and muonic hydrogen atoms that disagree significantly. The possible origins of the disagreement are the subject of much of this text. To summarize:
the possibilities are 
\begin{itemize}
\item The electronic hydrogen experiments are almost, but  not quite, as accurate as stated (Sect. 2.2).

\item The QED calculations are  almost, but  not quite, as accurate as stated (Sect. 4.2).

\item The two photon exchange term that depends on proton polarizability has not been correctly evaluated (Sect. 4.3).

\item The electron and muon really do have different interactions with the proton (Sect. 4.4), so that there is physics beyond the Standard Model.

\end{itemize}
None of these possibilities seem very likely, but all must be pursued. 

There are some tasks for theorists. While the results  of  many of the  QED calculations have been reconfirmed, it is still possible that something is missing in the theory. 
Current theories  of beyond the standard model physics are very primitive and need technical improvements before being realized as a complete gauge theory without anomalies.
Better treatments of the two-photon exchange term could be very useful.

Much additional work by experimentalists is necessary, and any future progress is contingent upon continuing a strong experimental effort in diverse directions.
The electronic hydrogen experiments need to be redone using modern techniques (Sect. 5.3).
A new high energy-low momentum transfer  electron scattering measurement could reach the accuracy of the spectroscopic determinations and provide an independent confirmation or repudiation of the radius obtained from electronic hydrogen spectroscopy, Sect. 5.1. The proposed  muon-proton scattering experiment (Sect. 5.2) would rule on theories of the two-photon exchange interaction and also on theories beyond the standard model.
Discovery of the new particles required by theories  beyond the standard model would make  such  theories believable.  Exotic atoms, in which muons are bound to  nuclei,
would also rule on all possible  theories, Sect. 5.4. Future work on positronium would put QED to detailed tests.

The proton is the only stable baryon and all of Life depends on it. It is important to understand the proton. Its radius should be a simple quantity to determine and understand, but this is not the case. Recent  experimental results have not been understood.  But future work will either reveal the  true value of this radius, a better understanding of its structure, or a very unexpected feature of its interactions.

\section{Acknowledgements}
The authors would like to thank the participants of the 2012 ECT*
Workshop on the Proton Radius Puzzle in Trento, Italy, for stimulating and
exciting discussions.
R.P.\ acknowledges support from the European Research Council (ERC), StG 279765 and thanks the members of the CREMA collaboration.
R.G.\ was supported by the US National Science Foundation grant PHY 09-69239. 
The work of G.A.M.\ is partially supported by USDOE  Grant No.
DE-FG02-97ER-41014. He would like to thank C.E.~Carlson, J.D.~Carroll, J.~Rafelski and A.W.~Thomas for useful discussions. 
K.P.\ was supported by NCN grant 2012/04/A/ST2/00105.


\end{document}